# Stress-Induced Transformations of Polarization Switching in CuInP$_2$S$_6$ Nanoparticles

Anna N. Morozovska[1*], Eugene A. Eliseev[2], Mykola E. Yelisieiev[3], Yulian M. Vysochanskii[4†], and Dean R. Evans[5‡]

[1] Institute of Physics, National Academy of Sciences of Ukraine,

Nauky Avenue, Bldg. 46, 03028 Kyiv, Ukraine

[2] Institute for Problems of Materials Science, National Academy of Sciences of Ukraine,

Krjijanovskogo Street, Bldg. 3, 03142 Kyiv, Ukraine

[3] Physics Faculty of Taras Shevchenko National University of Kyiv, Volodymyrska street 64, Kyiv,

01601, Ukraine

[4] Institute of Solid State Physics and Chemistry, Uzhhorod University,

88000 Uzhhorod, Ukraine

[5] Air Force Research Laboratory, Materials and Manufacturing Directorate,

Wright-Patterson Air Force Base, Ohio, 45433, USA

**Abstract**

Using the Landau-Ginzburg-Devonshire (LGD) approach, we study stress-induced transformations of polarization switching in ferrielectric CuInP$_2$S$_6$ nanoparticles for three different shapes: a disk, a sphere, and a needle. Semiconducting properties of a nanoparticle are modeled by a surface charge layer, whose effective screening length can be rather small due to the field-effect. We reveal a very strong and unusual influence of hydrostatic pressure on the appearance of polarization switching in CuInP$_2$S$_6$ nanoparticles, hysteresis loops shape, magnitude of the remanent polarization, and coercive fields, and explain the effects by the anomalous temperature dependence and "inverted" signs of CuInP$_2$S$_6$ linear and nonlinear electrostriction coupling coefficients. In particular, by varying the sign of the applied pressure (from tension to compression) and its magnitude (from zero to several hundreds of MPa), quasi-static hysteresis-less paraelectric curves can transform into double, triple, pinched, or single hysteresis loops. The shape of the quasi-static hysteresis loops is defined by specific static dependences of polarization on an applied electric field, referred to as "static curves". The structure of the static curves has very specific features for CuInP$_2$S$_6$, since its LGD potential is an 8-th order polynomial in the polarization powers. Due to the sufficiently wide temperature and pressure ranges of double, triple, and pinched hysteresis loop stability (at least in comparison with many other ferroelectrics), CuInP$_2$S$_6$ nanodisks can be of particular interest for applications in energy storage (in the region of double loops), CuInP$_2$S$_6$ nanospheres maybe suitable for dynamic random access multibit memory, and CuInP$_2$S$_6$ nanoneedles are promising for non-volatile multibit memory cells (in the regions of triple and pinched loops). The stress control of the polarization switching scenario allows the creation of advanced piezo-sensors based on CuInP$_2$S$_6$ nanocomposites.

---

[*] Corresponding author, e-mail: anna.n.morozovska@gmail.com

[†] Corresponding author, e-mail: vysochanskii@gmail.com

[‡] Corresponding author: dean.evans@afrl.af.mil

## 1. Introduction

Room-temperature uniaxial low-dimensional ferroics are very promising for advanced nanoscale non-volatile memory cells, information, and energy storage [1]. The feasible control of polarization switching by external stress, strain, or their gradients is especially useful to improve the memory performances of ferroics. That is why the stress-induced phase transitions [2] and strain engineering [3, 4] become especially important in nanoparticles and ultrathin films of Cu-based layered chalcogenides, CuInP$_2$(S,Se)$_6$ [5, 6, 7]; these materials are uniaxial ferroics [8, 9, 10], whose value for advanced applications is due to a possibility of the ferrielectricity and antiferrielectricity downscaling to the limit of a single layer [11]. Ferrielectricity, the equivalent of ferrimagnetism, can be termed as an antiferroelectric order, but with a switchable spontaneous polarization created by two sublattices with spontaneous dipole moments that are antiparallel and different in magnitude [12].

Despite the significant fundamental and practical interest in bulk [13] and nano-sized CuInP$_2$(S,Se)$_6$ [14], the influence of stress and strains on the switching of its spontaneous polarization has been very poorly studied from a theoretical perspective. There are a few mostly experimental works considering polarization switching in CuInP$_2$(S,Se)$_6$ ultrathin films and nanoflakes (see e.g., Refs. [3, 4] and references therein).

The spontaneous polarization of crystalline CuInP$_2$S$_6$ (**CIPS**) is directed normally to its structural layers as a result of antiparallel shifts of the Cu$^+$ and In$^{3+}$ cations from the middle of the layers [5, 13, 14]. Acentric positions of Cu$^+$ cations in the surrounding sulfur atoms octahedra are determined by the second order Jahn – Teller (SOJT) effect, which determines their two-well local potential with the Cu$^{up}$ and Cu$^{down}$ positions of the copper atoms in the ground state. The Cu$^+$ cations flip between Cu$^{up}$ and Cu$^{down}$ positions with a temperature increase and populate the positions with equal probability above the temperature of the structural transition from the polar ferrielectric (**FI**) to the nonpolar paraelectric (**PE**) phase. Complementary to the sublattice of the Cu$^+$ cations, the sublattice of In$^{3+}$ cations also plays an important role in the polar ordering of the CIPS crystal lattice. The indium cations with $4d^{10}3s^0p^0$ electronic configuration have some degree of covalence of their chemical bonding with the nearest sulfur atoms, which is determined by sp$^2$ hybridization because of their stereochemical activity, and form the three-well local potential with a stable central well and two metastable side wells [14]. The local potential determines the shift of the indium cations, which is opposite to the deviation of copper cations from the middle of CIPS structural layers.

Using pseudospin formalism, the polar ordering of CIPS can be described in the Ising model with spins $s = ½$ and $S = 1$ [15], and a mixed anisotropy of the local crystal field. Within the model, the spins $\vec{s}$ with projections + ½ and -½ can be associated with the local dipoles induced by Cu$^+$ cations and P$_2$S$_6$ anion complexes, and spins $\vec{S}$ with projections +1, 0, -1 can be related with the local dipoles induced by $In^{3+}$ cations and $P_2S_6$ anion complexes. In the mean-field approximation [16], the presence of two types of cationic sublattices in ferrielectrics is described by polar (dipolar) and antipolar (nonpolar) order parameters, $P$ and $A$, respectively. They formally correspond to projections $\vec{S} = \pm 1$ (in the polar state) and $\vec{S} = 0$ (in the nonpolar state) in the pseudospin formalism. Using the eight-power thermodynamic potential [2] in this case, we predicted a temperature – pressure phase diagram [17], containing the critical end point (**CEP**) where the first order PE – FI phase transition line terminates, and the bicritical end point (**BEP**) where the first order isostructural phase transition line between two ferrielectric states (**FI1** and **FI2**) with different amplitudes of spontaneous polarization terminates.

Using the eight-power free energy functional of polarization, Ishibashi and Hidaka [18] analyzed the phase diagram in a two-sublattice model with an asymmetric double-well potential. They calculated a phase diagram containing a triple point (analog of the CEP), an end point (analog of the BEP), and an isomorphous first order phase transition line between the triple point and the end point embedded in a polar region (analog of the first order isostructural transition between the FI2 and FI1 states in Ref.[17]). Their Kittel-type model considers polar ($P$) and antipolar ($A$) order parameters, expressed via the sublattices' polarizations $P_1$ and $P_2$, as $P = (P_1 + P_2)/2$ and $A = (P_1 - P_2)/2$, and includes a bilinear inter-sublattice coupling term, $gP_1P_2$. Ishibashi and Hidaka did not consider a switching of the spontaneous polarization.

Using the sixth power and fourth power free energy functional of polarization, Iwata and Ishibashi [19] consider the polarization switching in the first order ferroelectrics and second order antiferroelectrics, and found double hysteresis loops of polarization switching in a narrow temperature range above the Curie temperature (i.e., inside the PE phase). Toledano and Guennou [12] analyzed electric field-induced antiferroelectric – ferrielectric – ferroelectric transitions in the frame of the Kittel model with sixth order invariants in $P$ and in the presence of nonlinear coupling between $A$ and $P$. They defined the range of model parameters required for the appearance of double hysteresis loops. Balashova and Tagantsev [20] considered the fourth power free energy functional of a ferroic with a nonlinear coupling between the structural order parameter and the polarization, and reveal the existence of double hysteresis loops in the system. Recently, Lum et al. [21] developed the Kittel model by considering sixth order invariants and bilinear coupling between dipolar sublattices, and observed double hysteresis loops of polarization switching. Double and pinched hysteresis loops were observed in uniaxial ferroelectrics and were explained with the axial next nearest-neighbor Ising (ANNNI) model by Zamaraite et al. [22]. An early paper by Suzuki and Okada [23], which developed the Kittel model by taking the sixth order invariants, bilinear, and biquadratic nonlinear coupling terms, $gP_1P_2$ and $h(P_1P_2)^2$, described the phase diagram with paraelectric, polar (ferroelectric), semipolar (ferrielectric), and antipolar (antiferroelectric) phases. They predicted "anomalous" triple loops in the semipolar phase. It was mentioned, that the triple loops can be smeared at the rise of the coercive electric fields, because of the enhanced hysteresis effect. Pinched and triple loops were predicted in multiaxial ferroelectrics [24], and their importance for the creation of "symmetry-protected" ferroelectric multibit memory cells was emphasized.

Using the eighth-order Landau-Ginzburg-Devonshire (**LGD**) thermodynamic potential [2, 17] and finite element modeling (**FEM**), we study the stress-induced transformations of polarization switching in CIPS nanoparticles, whose shape varies, i.e., a prolate needle, an oblate disk, or a sphere. Much attention is paid to the analysis of the correlations between the peculiarities of spontaneous polarization switching by an electric field and the topology of the temperature – pressure phase diagram. We reveal relatively wide temperature and pressure ranges of double, triple, and pinched hysteresis loop stability (at least in comparison with many other ferroelectrics). The original part of this work contains the physical description of the problem and LGD approach (**Section II**), considers in detail the stress-induced transformations of a polarization switching scenario and hysteresis loop changes in the nanoparticles (**Section III.A** and **III.B**), and discusses possible applications of stressed CIPS nanoparticles in multibit memory cells (in the region of pinched and triple loops) and energy storage (in the region of double loops). **Section IV** provides conclusions of the results. **Supplementary Materials** contain a mathematical formulation of the problem, a table of material parameters, a description of methods, and numerical algorithms.

## II. Problem formulation

A CIPS nanoparticle, whose shape can be spherical, prolate ellipsoidal, or oblate ellipsoidal, is schematically shown in **Fig. 1(a)-(c)**, respectively. The spontaneous polarization, $P_s$, is directed along the CIPS polar axis "$X_3$", which coincides with the ellipsoid semi-axis $L$. Two other semi-axes in $X_1$ and $X_2$ directions are equal to $R$. As a matter of fact, $P_s$ is the total polarization of four possible polar-active sublattices (see Fig. 1 in Ref.[17]) in which individual dipole moments of the four Cu and four In atoms are antiparallel and different in magnitude in the absence of an external electric field and an applied pressure. The polarization can be determined from standard electrical measurements of the nanocomposite capacitance (e.g., as a remanent polarization of the hysteresis loop), as well as estimated by piezoelectric response microscopy of separated nanoflakes.

To analyze the polarization field dependence, $P_3(E_0)$, we consider a nanoparticle placed in a homogeneous quasi-static electric field $\vec{E}_0$ co-directed with its polar axis. The electric field $\vec{E}$ existing inside the nanoparticle is a superposition of the external $\vec{E}_0$ and the depolarization $\vec{E}_d$ fields created by the uncompensated bound charges (ferroelectric dipoles) near the particle surface and domain walls (if any exist).

Let us consider a CIPS nanoparticle placed in ambient conditions. In what follows we assume that the depolarization field is partially screened by the surface charge with density $\rho_s$ that covers the nanoparticle "core" as a screening "shell". The physical nature of the surface charge determines the dependence of $\rho_s$ on the acting electric field [25]. In what follows we account for the fact that an effective screening length, $\lambda$, associated with the density, $\rho_s$, is strongly field-dependent in ferrielectric-semiconductors, such as CIPS. It is the consequence of the depolarization field effect in ferroelectrics-semiconductors that causes the band bending near their surface. The band bending effect is weak for a small $E_0$, and therefore the density of free charges induced by $E_0$ is relatively small inside the core. At the same time, the "bare" (i.e., unscreened) depolarization field, $\vec{E}_d$, significantly exceeds the thermodynamic coercive field near the core surface. Strong band bending occurs in response to this bare field, and as a result the field-induced density of the screening charge becomes very high at the core surface, which decreases $\vec{E}_d$ in a self-consistent way. Since the charge density determines the value of $\lambda$, hereinafter we take into account that $\lambda$ is significantly different for the external and depolarizing fields, namely $\lambda(E_0) \gg \lambda(E_d)$.

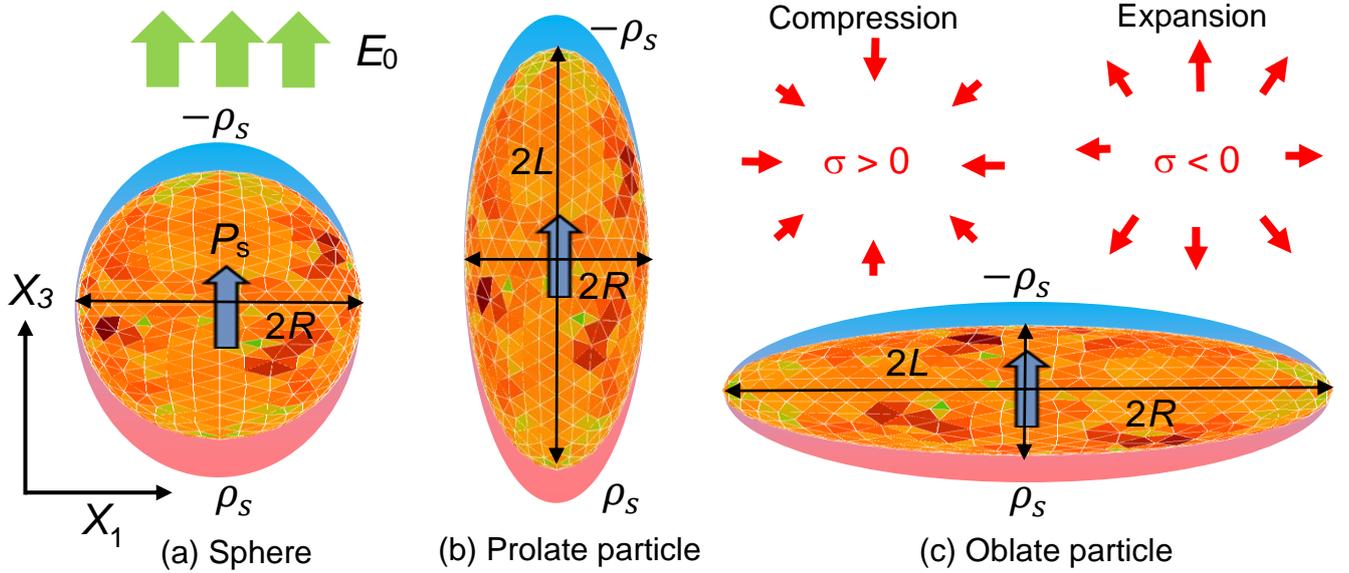

**FIGURE 1.** A CIPS nanoparticle whose shape can be spherical **(a)** with radius $R$, prolate ellipsoidal **(b),** or oblate ellipsoidal **(c)** with semi-axes $R$ and $L$. Mesh elements, used in FEM, are shown by light-green, orange, red, and dark-red colors. The shell of surface charge with density $\rho_s$ covers the nanoparticle core. The thick blue arrow shows the spontaneous polarization $P_s$ directed along the axis $X_3$; thinner red arrows illustrate the direction of the hydrostatic pressure application (compression, $\sigma > 0$, or expansion, $\sigma < 0$). The {$X_1$-$X_3$} cross-section of 3D ellipsoids is shown.

It has been shown in Refs.[2, 16, 17] that the density of the LGD free energy of CIPS has four potential wells at $\vec{E} = 0$. The free energy density includes the Landau-Devonshire expansion in even powers of the polarization $P_3$ (up to the eighth power), the Ginzburg gradient energy, and the elastic and electrostriction energies, which are listed in **Appendix A** of **Supplementary Materials** [26]. Polarization dynamics in an external field follows from the time-dependent LGD equation, which in turn follows from the minimization of the LGD free energy, and has the form:

$$\Gamma\frac{\partial P_3}{\partial t} + \left[\alpha - 2\sigma_i\left(Q_{i3} + W_{ij3}\sigma_j\right)\right]P_3 + (\beta - 4Z_{i33}\sigma_i)P_3^3 + \gamma P_3^5 + \delta P_3^7 - g_{33kl}\frac{\partial^2 P_3}{\partial x_k \partial x_l} = E_3. \quad (1)$$

$\Gamma$ is the Khalatnikov kinetic coefficient [27]. The coefficient $\alpha$ depends linearly on the temperature $T$, $\alpha(T) = \alpha_T(T - T_C)$, where $T_C$ is the Curie temperature of a bulk ferrielectric. The coefficients $\beta$, $\gamma$, and $\delta$ in Eq.(1) are temperature independent. The values $\sigma_i$ denote diagonal components of a stress tensor in the Voigt notation, and the subscripts $i, j = 1 - 6$. The values $Q_{i3}$, $Z_{i33}$, and $W_{ij3}$ denote the components of a single linear and two nonlinear electrostriction strain tensors in the Voigt notation, respectively [28, 29]. The values $g_{33kl}$ are polarization gradient coefficients in the matrix notation and the subscripts $k, l = 1 - 3$. The boundary condition for $P_3$ at the nanoparticle surface S is "natural", i.e., $g_{33kl}n_k\frac{\partial P_3}{\partial x_l}\big|_S = 0$, where $\vec{n}$ is the outer normal to the surface.

The value $E_3$ is an electric field component co-directed with the polarization $P_3$, which is a superposition of external and depolarization fields. In order to analyze a quasi-static polarization reversal, we assume that the period, $2\pi/\omega$, of the sinusoidal external field is very small in comparison with the Landau-Khalatnikov relaxation time, $\tau = \Gamma/|\alpha|$. The quasi-static field $E_3$ is related to the electric potential $\phi$ as $E_3 = -\frac{\partial \phi}{\partial x_3}$. The electric potential $\phi$ satisfies the Poisson equation inside the particle core,

$$\varepsilon_0 \varepsilon_b \Delta \phi = \frac{\partial P_3}{\partial x_3}. \quad (2)$$

The Laplace equation outside the screening shell is $\Delta \phi = 0$. The 3D Laplace operator is denoted by the symbol $\Delta$. Parameters $\varepsilon_0$ and $\varepsilon_b$ are the universal dielectric constant and the background dielectric permittivity [30] of the ferrielectric core, respectively. Equation (2) is supplemented by the condition of potential continuity at the particle surface, $(\phi_{ext} - \phi_{int})|_S = 0$. The boundary condition for the normal components of the electric displacement $\vec{D}$ is $\vec{n}(\vec{D}_{ext} - \vec{D}_{int})|_S = \rho_s$, where the surface charge density $\rho_s = -\varepsilon_0 \phi/\lambda$ and $\lambda$ is the above-mentioned effective screening length.

The values of $T_C$, $\alpha_T$, $\beta$, $\gamma$, $\delta$, $Q_{i3}$, $W_{ij3}$, and $Z_{i33}$ have been determined in Refs.[2, 16, 17] from the fitting of temperature-dependent experimental data for the dielectric permittivity [31, 32], spontaneous polarization [13, 33], and lattice constants [10] as a function of hydrostatic pressure. Elastic compliances $s_{ij}$ were calculated from ultrasound velocity measurements [34, 35]. The gradient coefficients $g_{33ij}$ were estimated from the width of domain walls. The CIPS parameters used in our calculations are listed in **Table SI** in **Appendix A.**

Due to the strong, negative, and temperature-dependent nonlinear electrostriction coupling coefficients (namely $Z_{i33} < 0$ and $W_{ij3} < 0$), and the "inverted" signs of the linear electrostriction coupling coefficients (namely $Q_{33} < 0$, $Q_{23} > 0$, and $Q_{13} > 0$) for CIPS, the expected pressure effect on the polarization switching is complex and unusual in comparison with many ferroelectrics where $Q_{33} > 0$, $Q_{23} < 0$, and $Q_{13} < 0$ [2]. We also would like to underline that below we consider only hydrostatic pressure, $\sigma_1 = \sigma_2 = \sigma_3 = \sigma$, since the case is the easiest to realize experimentally for an ensemble of nanoparticles. Note that the surface tension can renormalize the stresses as $\sigma_1 = \sigma - \frac{\mu}{R}$, $\sigma_2 = \sigma - \frac{\mu}{R}$, and $\sigma_3 = \sigma - \frac{\mu}{L}$ [36], where $\mu \cong (1-3)$ N/m is a relatively small surface tension coefficient [37, 38]. In order to focus on the influence of external pressure, we mostly neglect the surface tension in this work.

### III. Stress-induced transformations of polarization switching
### A. Finite Element Modeling of a polarization switching

We perform FEM in COMSOL@MultiPhysics software using electrostatics, solid mechanics, and general math (PDE toolbox) modules. FEM is performed for different nanoparticle sizes and shapes, discretization densities of the self-adaptive tetragonal mesh, and initial polarization distributions (e.g., randomly small fluctuations or poly-domain states) and their relaxation conditions (see details in **Appendix A**). Examples of mesh elements used in FEM are shown in **Fig. 1**. The smallest elements have a light-green color, larger elements are orange and red, and the largest have a dark-red color.

Typical quasi-static hysteresis loops, $P_3(E)$, calculated for a stress-free CIPS nanoparticle with radius $R = 6$ nm at 293 K, are shown in **Fig. 2(a)**. The loop shape strongly depends on the amplitude $V_0$ of the applied voltage (compare black, blue, red, and green loops). **Figure 2(a)** also shows that the polarization switching is mostly polydomain for a small $V_0$ value, and the corresponding loop has an unsaturated rhomb-like shape with small vertical steps (see black loop calculated for $V_0 = 0.02$ V). The total amount of domain walls and the fraction of polydomain states decrease with an increase in $V_0$, where the loop shape becomes more saturated and the vertical steps become much more pronounced (compare blue and red loops calculated for $V_0 = 0.1$ V and 0.2 V, respectively). Each step corresponds to a sharp change or "re-building" of the domain structure inside the nanoparticle. This re-building occurs at some critical voltage, as shown in the images 1-16 in **Fig. 2(b)**, which corresponds to the partially polydomain polarization switching in the red loop calculated for $V_0 = 0.2$ V. The sharp changes of the domain structure morphology mimic Barkhausen jumps [39] observed experimentally during the polarization reversal in many ferroelectrics. However, there are several principal differences between steps and Barkhausen jumps. For a specified particle size and temperature, the steps' position and height do not depend on the initial polarization distribution (i.e., on the sample pre-history) nor the number of voltage cycles; but it does depend on the average size of mesh elements. In contrast, Barkhausen jumps are determined by domain walls pinning by defects, and, as a rule, they become less visible or disappear after multiple voltage cycles.

It is interesting that the polydomain polarization states, shown in **Fig. 2(b)**, appear for a range of voltages when the voltage is increasing (see, e.g., points 2 - 8), but they are absent for the same magnitude of voltage when the voltage is decreasing (see, e.g., points 8 - 10). This is because once the applied field has created a stable single-domain state, it persists as a metastable state when the voltage is lowered from the maximal value to coercive value (e.g., to the point 10). Only below the coercive voltage is the metastable state destroyed by the electric field and the domain nucleation appears (see, e.g., points 11 - 16).

The polarization switching scenario exhibits a single-domain structure with a further increase of $V_0$, and the hysteresis loop acquires a saturated shape without any steps (see the dark-green loop calculated for $V_0 = 2$ V). The corresponding polarization distribution is quasi-uniform for the entire loop [see color images 1-6 in the plot **Fig. 2(c)**]. The coercive voltage corresponding to the dark-green loop (~0.14 V) is significantly higher than the coercive voltages corresponding to other loops (~0.01 – 0.02 V) in **Fig. 2(a)**. This is because a single-domain polarization switching occurs at the thermodynamic coercive field, which is significantly higher than the critical fields of domain wall motion. The shape of the single-domain loop is defined by the structure of the LGD potential only.

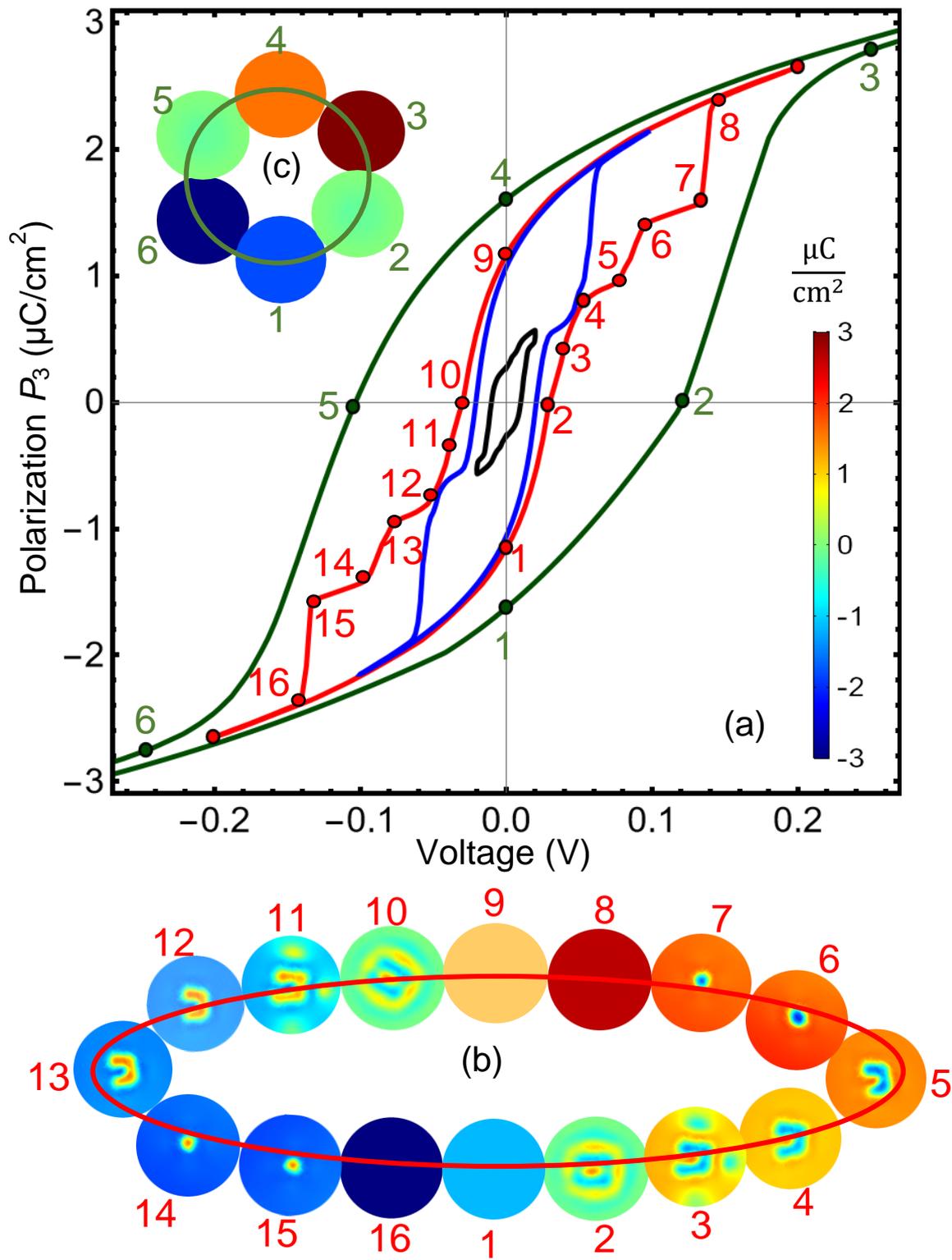

**FIGURE 2.** (a) Hysteresis loops, $P_3(E)$, calculated for a stress-free CIPS nanoparticle at different amplitudes of applied voltage: 0.02 V (black loop), 0.1 V (blue loop), 0.2 V (red loop), and 2 V (dark-green loop). Color images (b) and (c) are the polarization distribution in the equatorial cross-section of the particle perpendicular to its polar axis. Color images 2-7 and 10-15 in the plot (b) are the polydomain polarization states in the corresponding points on the red loop; and images 1, 8, 9, and 16 are the single-domain polarization states on the red loop. Color images 1-6 in the plot (c) are the single-domain polarization states in the points 1-6 on the dark-green loop. The color scale is the value of electric polarization in μC/cm². Nanoparticle radius $R = 6$ nm, $\lambda = 0.1$ nm, $T = 293$ K, and $\mu = 0$; CIPS parameters are listed in **Table SI.**

As it has been shown earlier [40], the domain structure appearance and its morphology in the ferrielectric state, as well as the critical sizes of the CIPS core corresponding to the FI-PE phase transition, depend strongly on the magnitude and anisotropy of the polarization gradient coefficients, $g_{33ij}$ (see Eq.(1)). In particular, a small increase in $g_{33ij}$ (e.g., above $10^{-9}$ J m³/C²) leads to a relatively small increase of the critical sizes (less than several nm), but it strongly suppresses the domain structure appearance in the FI state. For instance, in the case $g \cong 2 \times 10^{-9}$ J m³/C² (used in this work, see **Table SI**) and $\lambda = (0.1 - 0.5)$ nm, CIPS nanoparticles are mostly single-domain above the critical sizes (~ 5 nm) and PE below the critical sizes. Furthermore, for the case of natural boundary conditions used in this work, $g_{33ij} n_i \frac{\partial P_3}{\partial x_j} = 0$, polarization gradient effects can be neglected in the single-domain state. In order to study the role of anomalously strong nonlinear electrostriction coupling, the "inverted" signs of the linear electrostriction coupling, and the influence of the terms $\gamma P_3^5 + \delta P_3^7$ in Eq.(1) on the pressure effect of the polarization switching in CIPS nanoparticles, in what follows we limit our consideration to the single-domain polarization switching scenario.

### B. Pressure effect on the single-domain polarization switching scenario

The field dependence of a quasi-static single-domain polarization can be found from the following equation:

$$\Gamma \frac{\partial P_3}{\partial t} + \alpha^* P_3 + \beta^* P_3^3 + \gamma P_3^5 + \delta P_3^7 = E. \quad (3)$$

Here, $\beta^*(\sigma_i) = \beta - 4Z_{i33}\sigma_i$ and $E$ is the external field inside the core. The depolarization field, $E_d$, and stresses, $\sigma_i$, contribute to the "renormalization" of coefficient $\alpha(T)$, which becomes the temperature-, stress-, shape-, size-, and screening-dependent function $\alpha^*$ [17]:

$$\alpha^*(T, n_d, \sigma_i) = \alpha(T) + \frac{n_d}{\varepsilon_0[\varepsilon_b n_d + \varepsilon_s(1-n_d) + n_d(L/\lambda)]} - 2\sigma_i(Q_{i3} + W_{ij3}\sigma_j). \quad (4)$$

The derivation of the second term in Eq.(4) is given in Ref.[41]. Parameters $\varepsilon_b$ and $\varepsilon_s$ are the background dielectric permittivity [42] of the ferrielectric core and the relative dielectric permittivity of its screening shell or surrounding medium, respectively. Here $\lambda = \lambda(E_d)$, whose value can be rather small due to the surface band bending induced by the "bare" depolarization field $E_{d0}$ (see details after Eq.(S.5) in **Appendix A**). If we assume that $\lambda(E_0) \gg L$ and $\varepsilon_s \cong \varepsilon_b$, the estimate $E \approx \frac{\varepsilon_s E_0}{\varepsilon_b n_d + \varepsilon_s(1-n_d)} \cong E_0$ is valid. The dimensionless parameter $n_d$ is a shape-dependent depolarization factor introduced as [43]:

$$n_d(\xi) = \frac{1-\xi^2}{\xi^3}\left(\ln\sqrt{\frac{1+\xi}{1-\xi}} - \xi\right), \quad \xi = \sqrt{1 - \left(\frac{R}{L}\right)^2}. \quad (5)$$

Here $\xi$ is the eccentricity ratio of the ellipsoid, which depends on the dimensionless shape factor $\frac{R}{L}$. Both shape driven and size driven effects are intertwined in Eq.(4), because its second term, $\frac{n_d}{\varepsilon_0[\varepsilon_b n_d + \varepsilon_s(1-n_d) + n_d(L/\lambda)]}$, is a complex function of the shape factor $\frac{R}{L}$ (via the factor $n_d$) and the size $L$ in the polar direction (via the ratio $\frac{L}{\lambda}$).

The diagrams in **Fig. 3(a), 4(a),** and **5(a)** illustrate a typical influence of the hydrostatic pressure, $\sigma$, and temperature, $T$, on the value of spontaneous polarization, $P_s$, which corresponds to the absolute minimum of the LGD free energy for various phases (or states) of a CIPS nanodisk, nanosphere, and nanoneedle. All diagrams contain a large dark-violet region, where the PE phase with $P_s = 0$ is absolutely stable. As we have shown earlier [17], a part of the dark-violet region located between the dashed white curve and the sharp violet-red boundary contains the metastable FI phase, since CIPS undergoes a first order FI-PE phase transition at the sharp violet-red boundary. The magnitude of $P_s$ has a jump at the boundary. The PE phase becomes metastable in the region between the dotted white curve and the sharp violet-red boundary, where FI becomes absolutely stable. The dotted and dashed curves are the boundaries of the PE and FI phases absolute instability, and the PE and FI phases coexist between these curves. To underline this, we add thin red stripes (i.e., FI metastable and PE absolute stable regions) and violet stripes (i.e., FI absolute stable and PE metastable regions) in the region between dotted and dashed curves.

The intersection of the dashed and dotted curves is the CEP, denoted by a triangle in **Fig. 3(a)-5(a)**. Phase diagrams, shown in **Fig. 4(a)** and **5(a)**, also have the BEP marked by a white circle, which is positioned where the dashed and dot-dot-dashed curves intersect. The dot-dashed and dot-dot-dashed curves separate the region where the static dependences of polarization on the applied electric field (named "static curves") have four or six turning points; this number defines the structure of quasi-static hysteresis loops of polarization (as described in **Appendix B**). The static curves are shown as the thin black dashes inside the hysteresis loops in the right column (described in more detail below).

As we have shown earlier [17], the FI phase has two states. The large red region in **Fig. 3(a), 4(a),** and **5(a)** is the ferrielectric state 1 (**FI1**) with a relatively large $P_s$, and a small blue region is the ferrielectric state 2 (**FI2**) with a small $P_s$. The physical reason for the difference in $P_s$ in FI1 and FI2 states is the structure of four-well potential, schematically shown in inset to **Fig. 3(a)**. Here the red point in the deepest well of the red curve corresponds to a relatively large $P_s$ in the FI1 state, and the blue point in the deepest well of the blue curve corresponds to a relatively small $P_s$ in the FI2 state. Note that the color scale of $P_s$ corresponds to the deepest potential well of the LGD free energy.

Upon cooling the FI2 state transforms into the FI1 state. For **Fig. 4(a)** and **5(a)**, at temperatures below the BEP temperature and pressures smaller than the BEP pressure, a smooth isostructural transition from the FI2 to FI1 states occurs with a continuous increase of $P_s$. The polarization value, corresponding to the shallower well, is sensitive to the BEP, but it is not shown in **Figs. 4(a)-5(a).**

The diagrams in **Figs. 3(a)-5(a)** can be explained within the pseudospin formalism. The presence of the three-well local potential for the $In^{3+}$ cations in the CIPS lattice determines the observed peculiar shape of the $T - \sigma$ phase diagram with one PE and two ferrielectric (FI1 and FI2) states. The shape of the local potential determines the character of the phase transition from PE to FI states, either a first order transition with a step-like appearance of $P_s$ or a second order transition with a continuous increase of $P_s$. The stability of the local potential central well for the $In^{3+}$ cation determines the first order character of the PE - FI phase transition for $\sigma > 0$. The side well of the $In^{3+}$ cations' local potential is stabilized for $\sigma < 0$, and the second order PE-FI2 phase transition occurs.

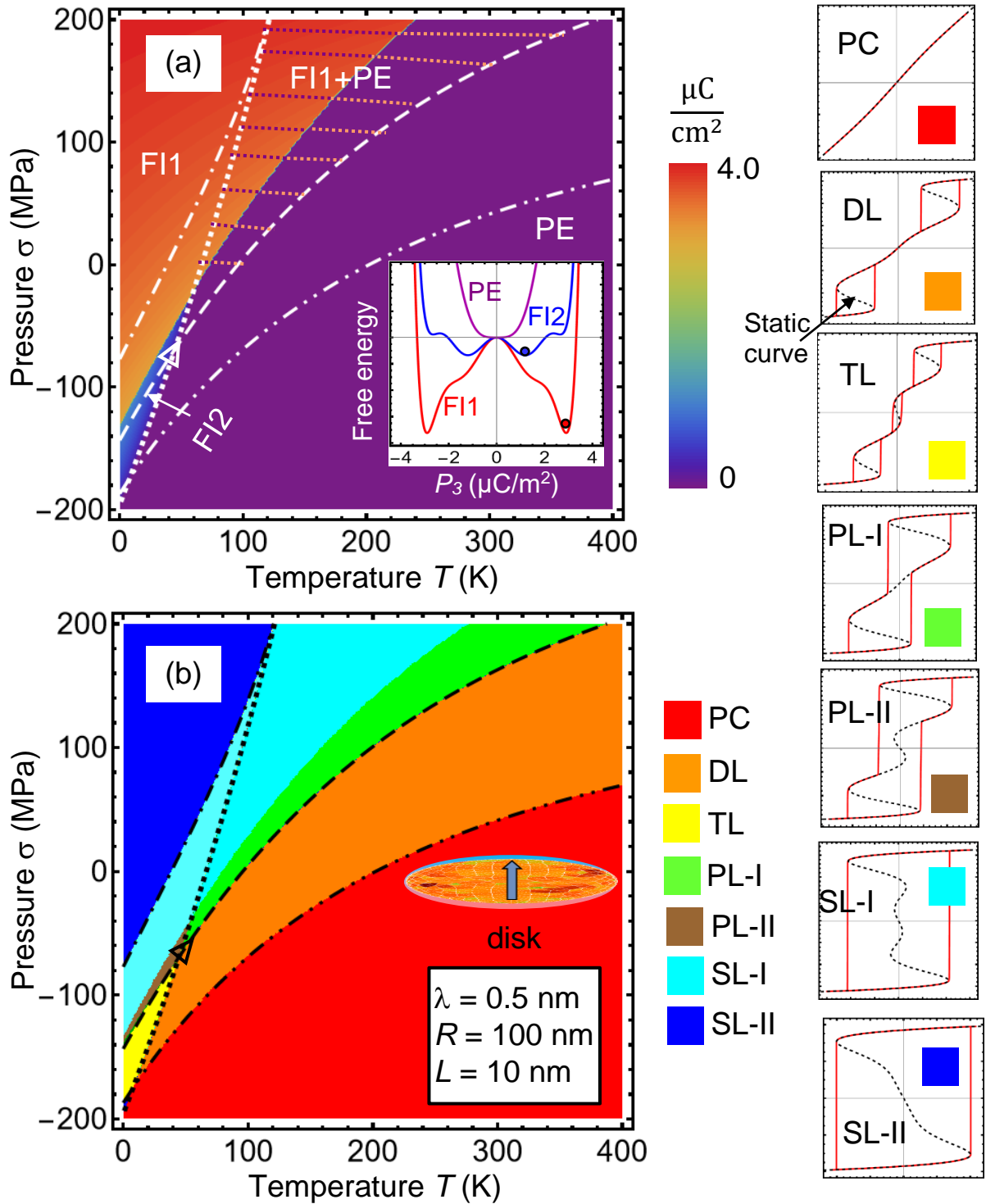

**FIGURE 3.** (a) The spontaneous polarization $P_s$, and (b) the shape of quasi-static hysteresis loops, $P_3(E)$, calculated as a function of temperature $T$ and pressure $\sigma$ for the stressed CIPS nanodisk with a radius $R = 100$ nm and a semi-height $L = 10$ nm. Color coding in the diagram (a) is the absolute value of $P_s$ in the deepest potential well of the LGD free energy. PE is the paraelectric phase, FI1 and FI2 are ferrielectric states; the CEP is marked by a triangle (see the explanation of abbreviations in the text). Color scale in the diagram (b): red is paraelectric curves (PC), orange is double loops (DL), yellow is triple loops (TL), light-green is pinched loops of the first type (PL-I), brown is pinched loops of the second type (PL-II), cyan is single loops of the first type (SL-I), and blue is single loops of the second type (SL-II). The static curves are shown as the thin black dashes inside the hysteresis loops in the right column. The screening length $\lambda = 0.5$ nm, surface tension coefficient $\mu = 0$; CIPS parameters are listed in **Table SI**.

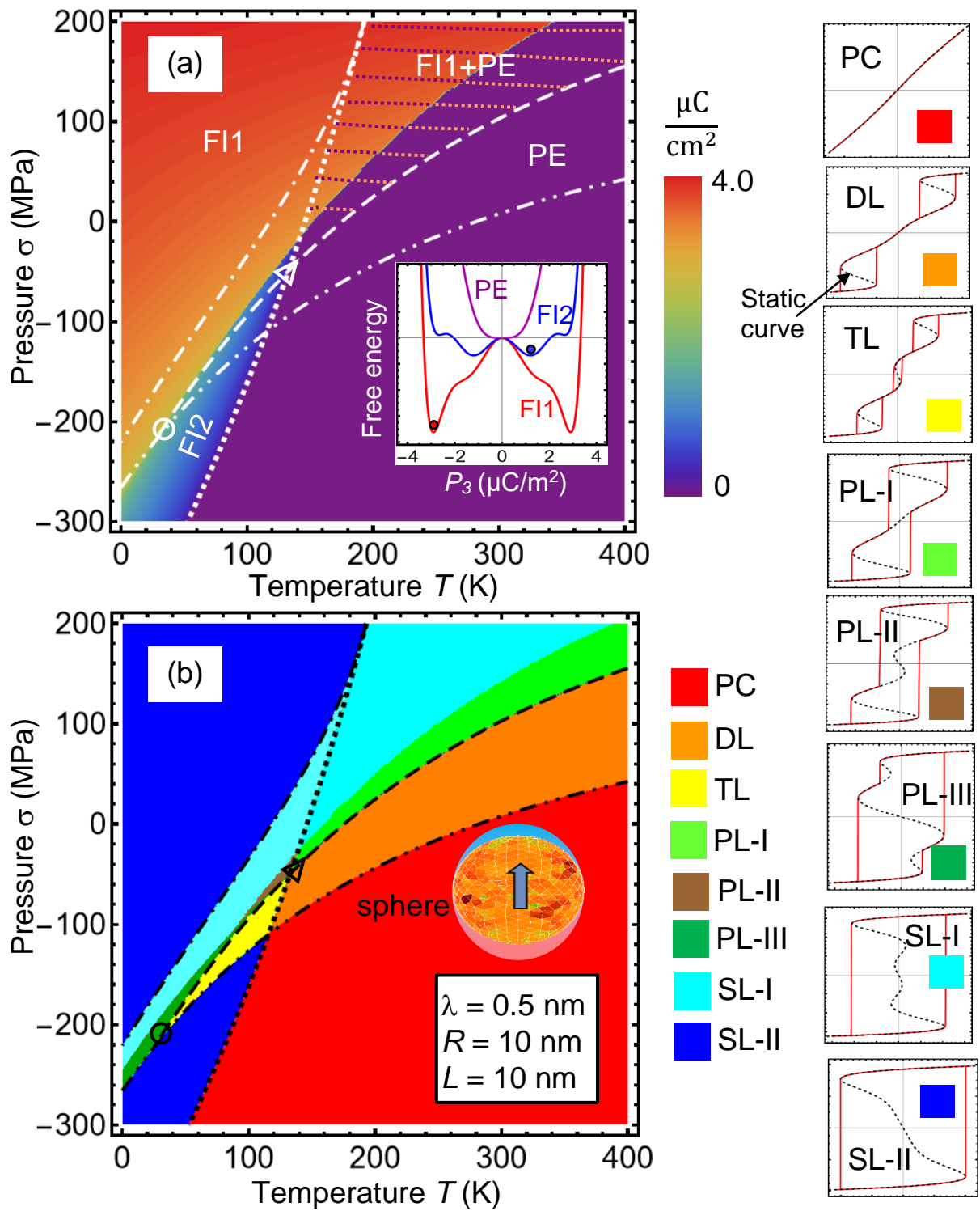

**FIGURE 4. (a)** The spontaneous polarization $P_s$, and **(b)** the shape of quasi-static hysteresis loops, $P_3(E)$, calculated as a function of temperature $T$ and pressure $\sigma$ for the stressed CIPS nanosphere with a radius $R = 10$ nm. Color scale in the diagram **(a)** is the absolute value of $P_s$ in the deepest potential well of the LGD free energy. PE is the paraelectric phase, FI1 and FI2 are ferrielectric states. The CEP is marked by a triangle, and the BEP is marked by a circle (see

the explanation of abbreviations in the text). Color coding in the diagram **(b)**: red is PC, orange is DL, yellow is TL, light-green is PL-I, brown is PL-II, dark-green is PL-III, cyan is SL-I, and blue is SL-II region (see the explanation of abbreviations in **Fig. 3**). The static curves are shown as the thin black dashes inside the hysteresis loops in the right column. Other parameters are the same as in **Fig. 3**.

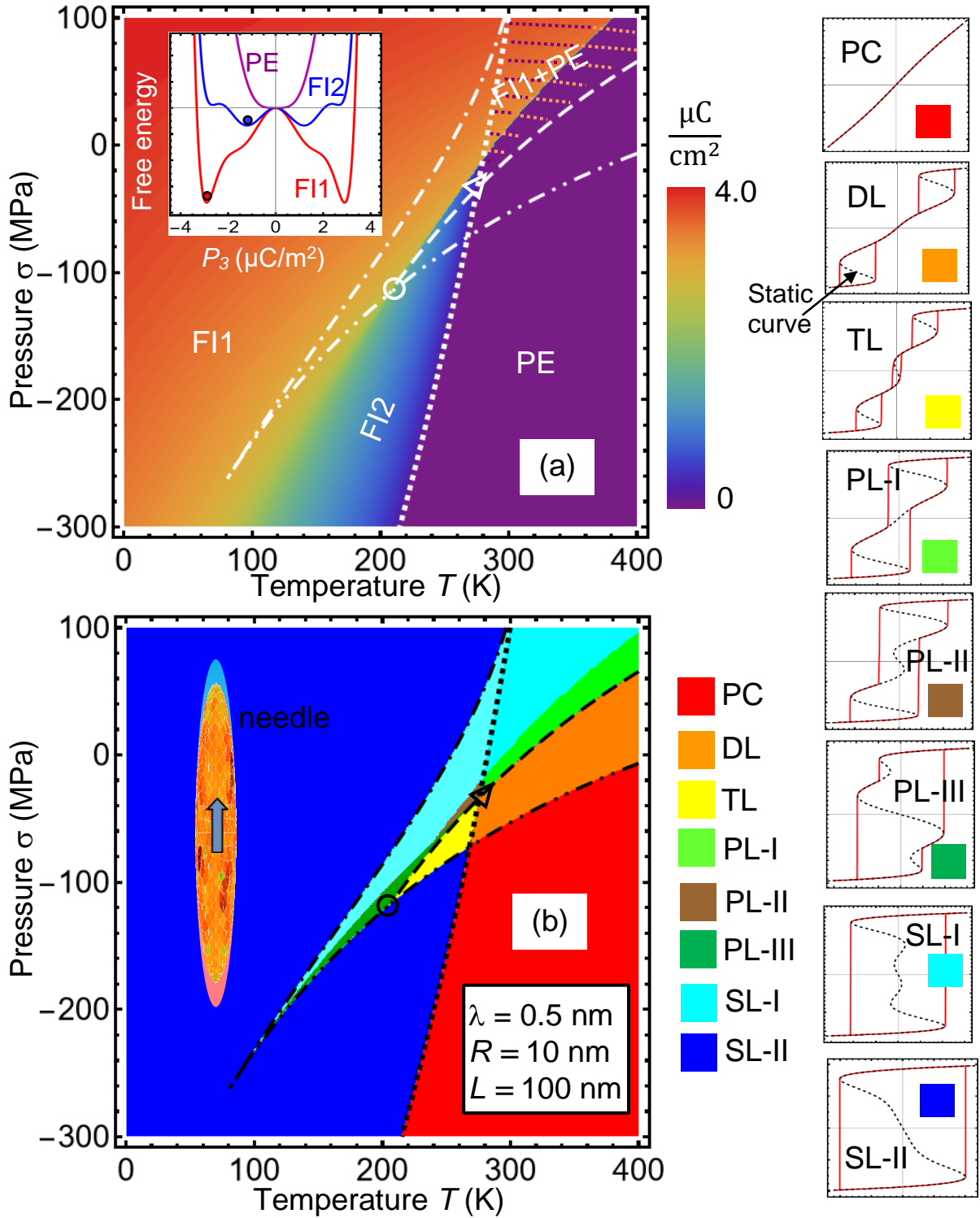

**FIGURE 5.** **(a)** The spontaneous polarization $P_s$, and **(b)** the shape of quasi-static hysteresis loops, $P_3(E)$, calculated as a function of temperature $T$ and pressure $\sigma$ for the stressed CIPS nanoneedle with a radius $R = 10$ nm and semi-

length $L = 100$ nm. Color scale in the diagram **(a)** is the absolute value of $P_S$ in the deepest potential well of the LGD free energy. PE is the paraelectric phase, FI1 and FI2 are ferrielectric states. The CEP is marked by a triangle, and the BEP is marked by a circle (see the explanation of abbreviations in the text). Color coding in the diagram **(b)**: red is PC, orange is DL, yellow is TL, light-green is PL-I, brown is PL-II, dark-green is PL-III, cyan is SL-I, and blue is SL-II region (see the explanation of abbreviations in **Fig. 3**). The static curves are shown as the thin black dashes inside the hysteresis loops in the right column. Other parameters are the same as in **Fig. 3**.

The behavior of the spontaneous polarization across the $T - \sigma$ diagram determines switching peculiarities of the polarization in different regions of the "loop" diagrams shown **Fig. 3(b), 4(b),** and **5(b)**. The diagrams in **Figs. 3(b)-5(b)** illustrate a typical influence of the hydrostatic pressure $\sigma$ and temperature $T$ on the shape of quasi-static hysteresis loops, $P_3(E)$, calculated for a CIPS nanodisk, nanosphere, and nanoneedle. The diagrams contain a red region of paraelectric curves (**PC**), an orange region of double loops (**DL**), a yellow region of triple loops (**TL**), a light-green region of pinched loops of the first type (**PL-I**), brown and dark-green regions of pinched loops of the second and third type (**PL-II** and **PL-III**), a cyan region of single loops of the first type (**SL-I**), and a blue region of single loops of the second type (**SL-II**). These different shapes of the hysteresis loops (red curves) and their corresponding static curves (thin black dashes inside the hysteresis loops) are shown in the right column; their detailed classification is given in **Appendix B**. The classification takes into account both the loop shape and the structure of the static curves.

The regions of PC and DL, separated by a dot-dot-dashed curve in **Figs. 3(b)-5(b)**, are located inside the PE phase region shown in **Figs. 3(a)-5(a)**. Indeed, the DL can be observed in the PE phase above the first order transition, as it was shown by Ishibashi [19]. These loops are related to a first order phase transition induced by the electric field.

The yellow region of TL in **Figs. 3(b)-5(b)** is located inside the blue region of FI2 phase shown in **Figs. 3(a)-5(a).** TLs fill the triangular region inside the dotted, dashed, and dot-dot-dashed curves in **Figs. 3(b)-5(b)**. The CEP and BEP are 2 vertices of the triangle. As it follows from the structure of the static curve, a TL can be imagined as the result of superposition of a "central" single loop (corresponding to the switching of the "small" polarization in the FI2 state) and a double loop (corresponding to the switching of a "larger" polarization in the FI2 state). The coexistence of these two polarizations in the four-well FI2 state determines the stability of TL in the yellow triangular region in **Figs. 3(b)-5(b)**.

The shape and transformation of the pinched loops in the light-green (PL-I), brown (PL-II), and dark-green (PL-III) regions in **Figs. 3(b)-5(b)** are also determined by the temperature and pressure evolution of superposed single and double loops. The region of PL-I is located above the dashed curve in the PE and FI1 coexisting region. Thin strips of PL-II and PL-III regions are located near the diffuse boundary between FI2 and FI1 states. The cyan triangular region of SL-I partially fills the region between the dot-dashed, dot-dot-dashed, and dashed curves in **Figs. 3(b)-5(b)**. The SL-I region is located inside the FI1 state in **Figs. 3(a)-5(a)**. The SL-I region is also stable in the PE and FI1 coexistence region. The large blue region of SL-II in **Figs. 3(b)-5(b)** is located inside the FI1 and FI2 states' regions of absolute stability in **Figs. 3(a)-5(a)**. The absolute stability of the FI states appears when the four-well potential transforms into the two-well potential, shown by red curves with a red filled circle in the insets in **Figs. 3(a)-5(a).**

A common feature of **Figs. 3-5** is the strong anomalous influence of the hydrostatic pressure on the appearance of polarization switching in CIPS nanoparticles, i.e., the unusual shape hysteresis loops and particularly the existence of the wide regions of DL, PL-I, and TL. The strong increase of the polarization switching region for $\sigma > 0$ and the increase of the hysteresis-less region for $\sigma < 0$, is caused by the anomalous temperature-dependence and "inverted" sign of the CIPS linear and nonlinear electrostriction coupling coefficients. The unusual shape of the quasi-static single-domain hysteresis loops, including the DL, PL, and TL, is defined by the specific structure of the static curves, which is determined by the 8-th order LGD potential. The above-mentioned properties of polarization switching are common for CIPS nanodisks, nanospheres, and nanoneedles, but there are some distinct features which are analyzed below.

In the case of CIPS nanodisks, shown in **Fig. 3(a)**, the region of the FI1 state is relatively small and strongly decreases with a temperature increase. Under the expansion pressure of -40 MPa (or greater in magnitude) a very small region of the FI2 state appears and continuously transforms into the PE phase. The FI1-PE and FE1-FE2 phase boundaries are sharp. The PC region prevails in tension-stressed ($\sigma < 0$) nanodisks; the wide SL-II, PL-I, and DL regions prevail for compression-stressed ($\sigma > 0$) nanodisks in the diagram in **Fig. 3(b)**. Four types of loops, namely the PL-II, TL, DL, and PL-I, coexist in the CEP. The PL-II region is almost absent for disks, occupying a very small region at low temperatures and $\sigma < 0$. The PL-III region is absent for disks, at least for the considered parameters, and a small region of TL is located at low temperatures (below 40 K). Since sizeable regions of PL-III and TL loops are required for nonvolatile multibit memory cells [24], CIPS nanodisks do not look promising for advanced memories. Since the wide region of DL are important for energy storage, CIPS nanodisks may be quite suitable for this application.

In the case of CIPS nanospheres, shown in **Fig. 4(a)**, the region of the FI1 state is much larger compared to the case of nanodisks, shown in **Fig. 3(a)**. Under the expansion pressure of -40 MPa (or greater in magnitude) a relatively wide region of the FI2 state appears and continuously transforms into the PE phase. The first order FI1-PE phase boundary is sharp, and the FI1-FI2 boundary is sharp above and diffuse (i.e., second order) below the BEP. Comparing **Fig. 4(b)** with **3(b)**, one can see that a thin region of PL-III appears for $\sigma < 0$ in the temperature range from 0 to 100 K. The dominant regions for the case of the nanodisk [**Fig. 3(b)**] for high temperatures and $\sigma < 0$ is PC and for temperatures lower than 250 K and $\sigma > 0$ is SL-I and SL-II; CEP is present and BEP is absent. For nanospheres [**Fig. 4(b)**] there are two special points, CEP and BEP, located in the middle of the loop diagram. The SL-II, PL-III, and TL loops coexist in the BEP. The size of the SL region increases compared to the case of the nanodisks. The temperature range of the PL-III and TL regions stability is from 0 to 120 K. The DL are stable from 100 to 400 K. Since relatively large regions of the pressure-induced PL-III, TL, and DL loops are predicted for stressed CIPS nanospheres, they can be promising candidates for energy storage (from 100 to 400 K) and multibit memory cells (from 0 to 120 K).

The diagrams of CIPS nanoneedles, shown in **Fig. 5**, look similar to the diagrams of nanospheres, shown in **Fig. 4**. However, the transition region between FI1 and FI2 states in **Fig. 5(a)** is wider and shifted toward higher temperatures and pressures in comparison with **Fig. 4(a)**. Due to the size effect, the SL-II region is larger, and the sizes of DL, PL-I, PL-II, PL-III, and TL regions are slightly smaller for nanoneedles than for nanospheres and shifted to significantly higher temperatures [compare **Fig. 4(b)** and **Fig. 5(b)**]. Indeed, the depolarization field contribution, given by the term $\frac{n_d}{\varepsilon_0[\varepsilon_b n_d + \varepsilon_s(1-n_d) + n_d(L/\lambda)]}$ in Eq.(4), is smaller for nanoneedles with a semi-length $L = 100$ nm than for the nanospheres with a radius $R = 10$ nm. The CEP and BEP are shifted to higher temperatures in comparison to the previous case. Since the temperature range of the TL region stability reaches 260 K at $\mu = 0$, which is relatively close to room temperature, CIPS nanoneedles are more promising candidates for multibit memory cells than nanospheres, where the TL region stability reaches 120 K at $\mu = 0$.

Next, we analyze the pressure influence on polarization switching at room temperature, since the temperature is a critical parameter for most applications. It is seen from **Fig. 6**, calculated for a CIPS nanodisk, nanosphere, and nanoneedle, that by varying the sign of applied pressure (from expansion to compression) and its magnitude (from zero to several hundreds of MPa), a quasi-static hysteresis-less paraelectric curve can transform into a double, pinched, or single hysteresis loop. The above-mentioned properties of polarization switching are qualitatively similar for CIPS nanodisks, nanospheres, and nanoneedles at room temperature, but there are some quantitative differences which are analyzed below.

For CIPS nanodisks, the pressure-induced transition from double to single hysteresis loops occurs at relatively high compression pressures [see **Fig. 6(a)-(c)** and **Fig. S9** in **Appendix C**]. As the pressure increases from 70 MPa to 170 MPa, small double loops grow and become more pronounced, finally merging together into a pinched loop of the first type (PL-I). A further increase of pressure from 170 MPa to 210 MPa leads to a gradual growth of the pinched section width, and the pinched loop finally transforms into a single loop of the second type (PL-II). The two coercive fields of the double loops vary in range (0 – 60) mV/nm and decrease with an increase in pressure; the coercive field of pinched and single loops increases from 20 mV/nm to 60 mV/nm with an increase in pressure.

For CIPS nanospheres, the pressure-induced transition from double to pinched hysteresis loops occurs under a pressure increase from 20 MPa to 100 MPa [see **Fig. 6(d)-(f)** and **Fig. S10** in **Appendix C**]. A further increase of pressure from 100 MPa to 140 MPa leads to a gradual growth of the pinched section, and the pinched loop of the third type (PL-III) eventually transforms into a single loop of the first type (SL-I). The coercive fields for nanospheres are one and a half times smaller than the fields for nanodisks. For CIPS nanoneedles, the pressure-induced transition from double hysteresis loops to pinched loops, and then to single loops, occurs within a much narrower pressure range, from -40 MPa to 10 MPa [see **Fig. 6(g)-(i)** and **Fig. S11** in **Appendix C**]. The coercive fields for nanoneedles are approximately two-times smaller than the fields for nanospheres.

**Figure 7** shows the onset and temperature evolution of stress-induced PL-III and TL hysteresis loops in CIPS nanoneedles in the temperature range (265 – 300) K. Under zero surface tension ($\mu = 0$), the TL appears from the PL-III in the temperature range (265 – 270) K for a pressure $\sigma \cong -50$ MPa, and gradually transforms into a DL near 280 K (see **Figs. 7a-7c**). SL-I and PL-III are stable for stress-free nanoneedles (see **Figs. 9d** and **Fig. S12** in **Appendix C**). Negative surface tensions, $\mu \cong -(0.7 - 1)$ N/m, induce a TL around 290 K for zero external pressure, $\sigma = 0$ (see **Figs. 7d-6f**). Hence it is possible to shift the $T - \sigma$ range of the PL-III and TL stability to zero stress and room temperature by a small negative surface tension, which shows the CIPS nanoneedles' potential for nonvolatile and/or dynamic multibit memory cells.

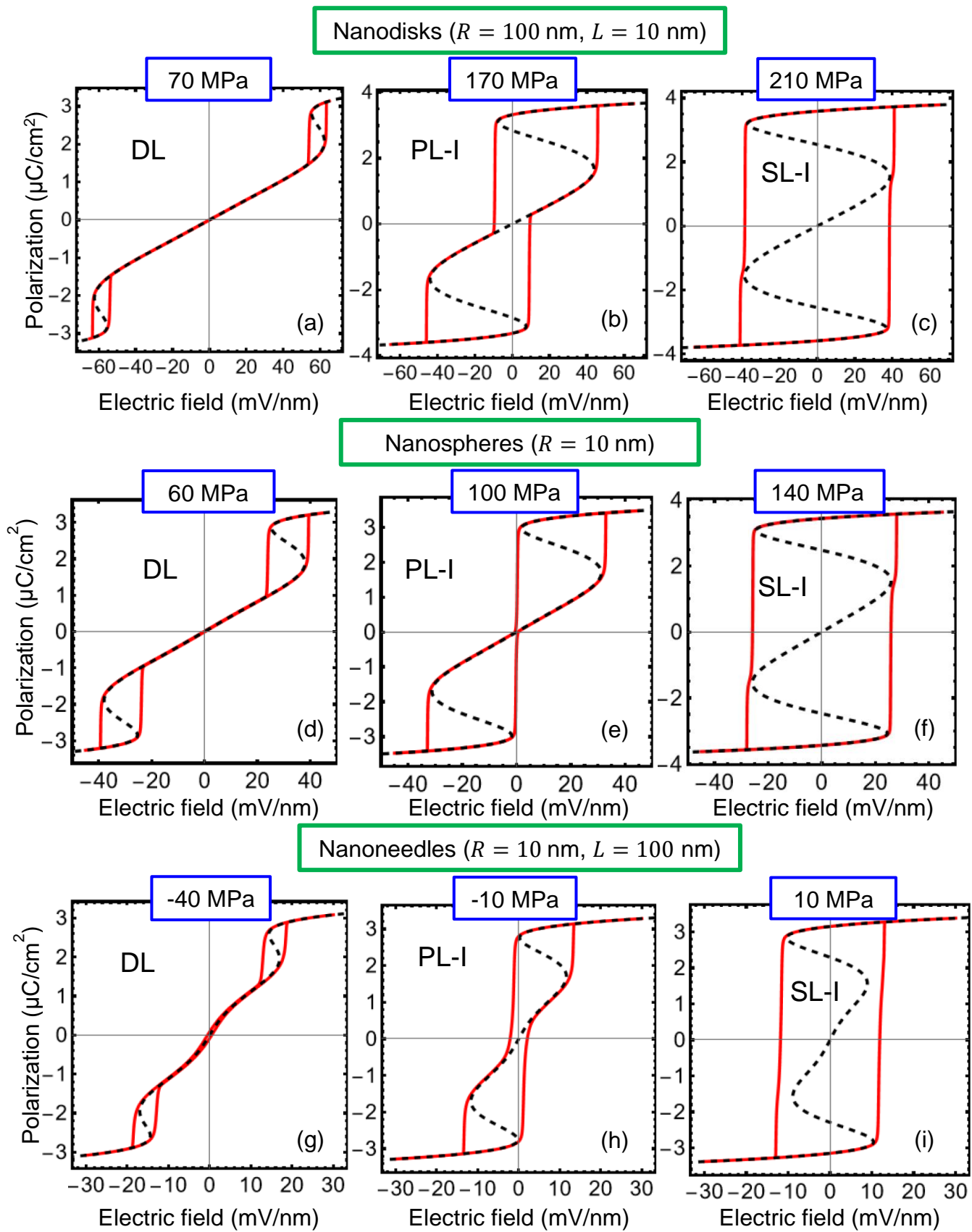

**FIGURE 6**. Electric field dependence of the polarization $P_3$, calculated for CIPS nanodisks with $R = 100$ nm, $L = 10$ nm **(a-c)**, nanospheres with $R = 10$ nm **(d-f)**, and nanoneedles with $R = 10$ nm, $L = 100$ nm **(g-i)**, under different hydrostatic pressures: $\sigma = 70$ MPa **(a)**, 170 MPa **(b)**, 210 MPa **(c)**, 60 MPa **(d)**, 100 MPa **(e)**, 140 MPa **(f)**,

-40 MPa **(g)**, -10 **(h)**, and 10 MPa **(i).** Black dashed curves are static dependences and red solid loops are quasi-static hysteresis loops. Temperature $T$ =300 K, screening length $\lambda = 0.5$ nm; other parameters are the same as in **Fig. 3**.

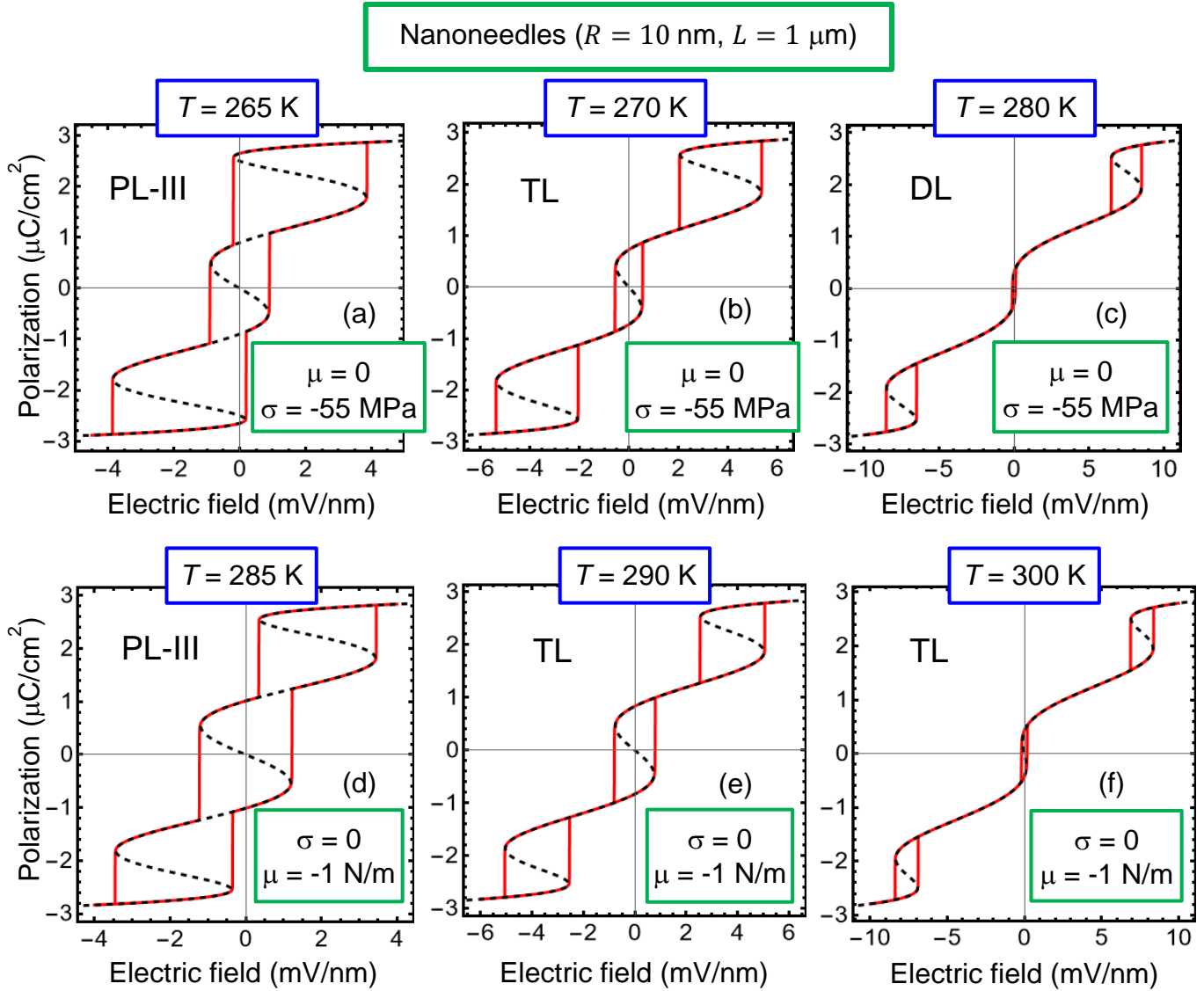

**FIGURE 7.** Electric field dependence of the polarization $P_3$, calculated for CIPS nanoneedles with sizes $R = 10$ nm and $L = 1$ μm. Temperature, $T$, hydrostatic pressure, $\sigma$, and surface tension coefficient, $\mu$, are listed in the legends. Black dashed curves are static dependences and red solid loops are quasi-static hysteresis loops. The screening length $\lambda = 0.5$ nm; other parameters are the same as in **Fig. 3**.

### C. Possible applications

We predict that compressed CIPS nanodisks reveal wide temperature and pressure ranges of DL stability in comparison with multiaxial perovskite ferroelectrics, such as BaTiO$_3$ and PbTiO$_3$ single-crystals. In particular, DL are stable in the temperature range (100 – 400) K at pressures (0 – 70) MPa [see **Fig. 3(b)**]. At zero pressure, the width of the DL region is 100 K, while the width reaches 200 K at $\sigma = 40$ MPa. Hence, compression-stressed CIPS nanodisks as energy storage nanomaterials can be competitive with classical antiferroelectrics, such as PbZrO$_3$ thin films [44]. Indeed, CIPS nanodisks' in-field polarization can reach 4 μC/cm$^2$, and their thermodynamic coercive field varies in the range (5 - 50) mV/nm. PbZrO$_3$ thin films have much higher in-field polarization (~40 μC/cm$^2$) and also much higher thermodynamic coercive fields (~500 mV/nm). Of course, the stored energy in a CIPS nanoflake [proportional to the area above the DL, see green region in **Fig. 8(a)**] is much smaller than in a PbZrO$_3$ thin film, but the losses and writing voltage are also much smaller for CIPS. The region of DL stability for stressed CIPS nanospheres and nanoneedles is smaller than for nanodisks [see **Fig. 4(b)** and **5(b)**], but it is still relatively wide, e.g., 150 K at zero pressure, which is a large value in comparison with most ferroelectrics where the width does not exceed (10 – 50) K.

Unexpectedly, we predict that stressed CIPS nanospheres and nanoneedles reveal sizeable temperature and pressure ranges of TL and PL-III stability [see **Fig. 4(b)** and **5(b)**], which are very rare in ferroelectrics and antiferroelectrics. These types of loops can be used for dynamic and/or nonvolatile multibit memory cells, as proposed in Ref.[24] and schematically shown in **Fig. 8(b)** and **8(c)**, respectively. The stability of PL-III and TL at zero external stress, room temperature, and relatively small negative surface tension $\mu$ makes nanocomposites with CIPS nanoneedles promising for use in dynamic and/or nonvolatile multibit memory cells.

The pressure-induced transition of the hysteresis loops in composites with CIPS nanoparticles can be used in precise pressure sensors for high-pressure applications, such as air compressors or shockwave detectors. The pressure-induced transitions of the polar state and switching scenario in CIPS nanoparticles occur at relatively low pressures (e.g., several MPa) in comparison with other ferroelectrics (~ GPa), which is quite reasonable for practical applications. This can be useful for high-pressure piezo-sensors, because this could lead to higher precision of pressure measurements.

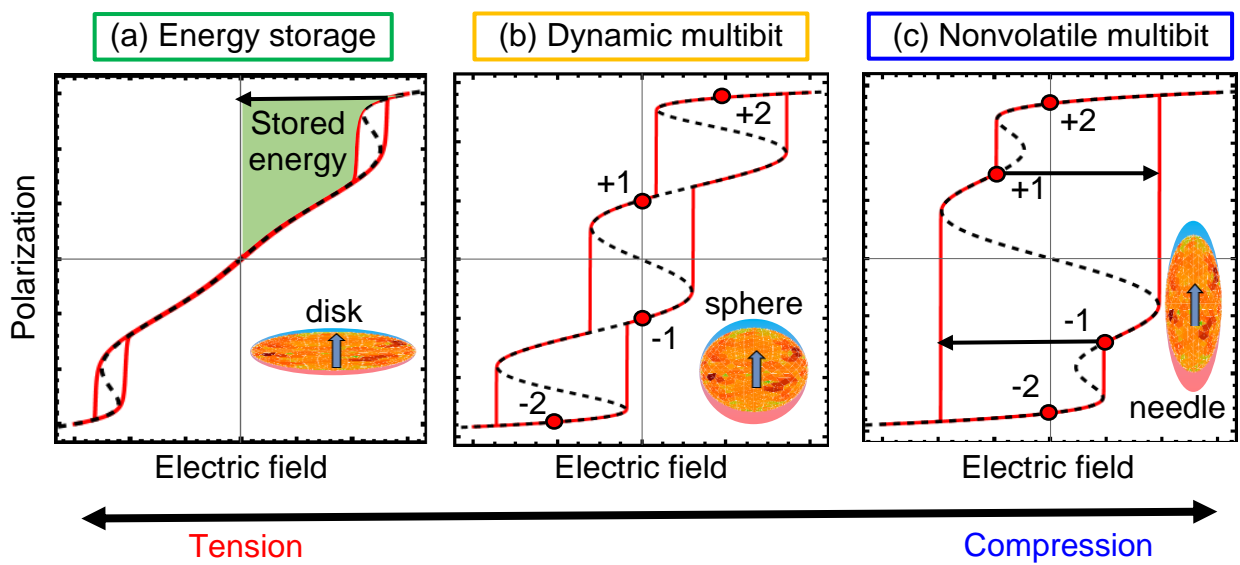

**FIGURE 8.** CIPS nanoparticles for energy storage **(a)**; and dynamic **(b)** and nonvolatile **(c)** multi-bits with memory states ±1 and ±2.

# IV. Conclusions

We reveal an unusually strong influence of hydrostatic pressure on the appearance of polarization switching in CIPS nanoparticles, hysteresis loops shape, magnitude of the remanent polarization, and coercive fields, which is explained by the effect of the anomalous temperature-dependence and "inverted" sign of CIPS linear and nonlinear electrostriction coupling coefficients. In particular, by varying the sign of the applied pressure (from expansion to compression) and increasing its magnitude (from zero to several hundreds of MPa), a quasi-static hysteresis-less paraelectric dependence can transform into a double, triple, pinched, or single hysteresis loop. The form of quasi-static hysteresis loops is defined by specific static dependences of polarization on applied electric field, i.e., "static curves". The structure of the static curves has very specific features for CIPS, since its LGD potential is an 8-th order polynomial in the polarization powers.

Due to the sufficiently wide temperature and pressure ranges of double, triple, pinched, and single hysteresis loop stability (at least in comparison with many other ferroelectric materials), stressed CIPS nanodisks can be of particular interest for applications in energy storage (in the region of double loops), and CIPS nanoneedles and nanospheres can be used in nonvolatile and/or dynamic multibit memory cells (in the region of PL-III and TL loops). Since the temperature range of the TL region stability is relatively low for nanospheres and close to room temperature for CIPS nanoneedles, the nanoneedles are more promising candidates for multibit memory cells than nanospheres. The stress control of the polarization switching scenario allows the creation of advanced piezo-sensors based on nanocomposites with CIPS nanospheres.

**Acknowledgements.** A.N.M. acknowledges EOARD project 9IOE063 and related STCU partner project P751a. E.A.E. acknowledges support of the National Academy of Sciences of Ukraine.

**Authors' contribution.** The research idea belongs to A.N.M. and Yu.M.V. A.N.M. formulated the problem, performed analytical calculations, analyzed results, and wrote the manuscript draft. E.A.E. and M.E.Y. wrote codes; A.N.M and M.E.Y. prepared figures and Suppl. Materials. Yu.M.V. and D.R.E. worked on the results explanation and manuscript improvement. All co-authors discussed the obtained results.

# References

[1]     J. S. Meena, S. M. Sze, U. Chand, and T.-Y. Tseng, Overview of emerging nonvolatile memory technologies, Nanoscale Res. Lett. **9**, 526 (2014).

[2]     A. N. Morozovska, E. A. Eliseev, S. V. Kalinin, Y. M. Vysochanskii, and Petro Maksymovych. Stress-Induced Phase Transitions in Nanoscale CuInP$_2$S$_6$. Phys. Rev. B **104**, 054102 (2021).

[3] C. Chen, H. Liu, Q. Lai, X. Mao, J. Fu, Z. Fu, and H. Zeng. Large-Scale Domain Engineering in Two-Dimensional Ferroelectric CuInP$_2$S$_6$ via Giant Flexoelectric Effect. Nano Letters **22**, 3275 (2022).

[4] W. Yang, S. Chen, X. Ding, J. Sun, and J. Deng. Reducing Threshold of Ferroelectric Domain Switching in Ultrathin Two-Dimensional CuInP$_2$S$_6$ Ferroelectrics via Electrical−Mechanical Coupling. J. Phys. Chem. Lett. **14**, 379 (2023).

[5] X. Bourdon, V. Maisonneuve, V.B. Cajipe, C. Payen, and J.E. Fischer. Copper sublattice ordering in layered CuMP$_2$Se$_6$ (M=In, Cr). J. All. Comp. **283**, 122 (1999).

[6] A. Belianinov, Q. He, A. Dziaugys, P. Maksymovych, E. Eliseev, A. Borisevich, A. Morozovska, J. Banys, Y. Vysochanskii, and S. V. Kalinin, CuInP$_2$S$_6$ Room Temperature Layered Ferroelectric. Nano Lett. **15**, 3808 (2015).

[7] M. A. Susner, M. Chyasnavichyus, M. A. McGuire, P. Ganesh, and P. Maksymovych. Metal Thio- and Selenophosphates as Multifunctional van der Waals Layered Materials. Advanced Materials **29**, 1602852 (2017).

[8] M. Wu, and P. Jena, The rise of two-dimensional van der Waals ferroelectrics. Wiley Interdisciplinary Reviews: Computational Molecular Science **8**, e1365 (2018).

[9] F. Liu, L. You, K. L. Seyler, X. Li, P. Yu, J. Lin, X. Wang, J. Zhou, H. Wang, H. He, S.T. Pantelides, W. Zhou, P. Sharma, X. Xu, P.M. Ajayan, J. Wang and Z. Liu, Room-temperature ferroelectricity in CuInP$_2$S$_6$ ultrathin flakes. Nature Communications **7**, art. num. 12357 (2016).

[10] M. A. Susner, M. Chyasnavichyus, A. A. Puretzky, Q. He, B. S. Conner, Y. Ren, D. A. Cullen et al. Cation–Eutectic Transition via Sublattice Melting in CuInP$_2$S$_6$/In$_{4/3}$P$_2$S$_6$ van der Waals Layered Crystals. ACS Nano **11**, 7060 (2017).

[11] M. Osada, and T. Sasaki. The rise of 2D dielectrics/ferroelectrics. APL Materials **7**, 120902 (2019).

[12] P. Toledano and M. Guennou. Theory of antiferroelectric phase transitions. Phys. Rev. B **94**, 014107 (2016).

[13] V. Maisonneuve, V. B. Cajipe, A. Simon, R. Von Der Muhll, and J. Ravez. Ferrielectric ordering in lamellar CuInP$_2$S$_6$. Phys. Rev. B **56**, 10860 (1997).

[14] Ju. Banys, A. Dziaugys, K. E. Glukhov, A. N. Morozovska, N. V. Morozovsky, Yu. M. Vysochanskii. Van der Waals Ferroelectrics: Properties and Device Applications of Phosphorous Chalcogenides (John Wiley & Sons, Weinheim 2022) 400 pp. ISBN: 978-3-527-35034-6

[15] W. Selke and J. Oitmaa, Monte Carlo study of mixed-spin *S = (1/2, 1)* Ising ferrimagnets, J. Phys.: Condensed Matter, **22**, 076004 (2010).

[16] A. N. Morozovska, E. A. Eliseev, K. Kelley, Yu. M. Vysochanskii, S. V. Kalinin, and P. Maksymovych. Phenomenological description of bright domain walls in ferroelectric-antiferroelectric layered chalcogenides. Phys. Rev. B **102**, 174108 (2020).

[17] A. N. Morozovska, E. A. Eliseev, Yu. M. Vysochanskii, V. V. Khist, and D. R. Evans. Screening-Induced Phase Transitions in Core-Shell Ferroic Nanoparticles. Phys. Rev. Materials **6**, 124411 (2022).

[18] Y Ishibashi, Y Hidaka. On an isomorphous transition, J. Phys. Soc. Jap. **60(5),** 1634-1637 (1991).

[19] M. Iwata, Y. Ishibashi. Phenomenological theory in dielectric tunable materials with the tristable states, Ferroelectrics, 503:1, 7-14 (2016), DOI: 10.1080/00150193.2016.1216700

[20] E.V. Balashova and A.K. Tagantsev, Polarization response of crystals with structural and ferroelectric instabilities, Phys. Rev. B **48,** 9979 (1993).

[21]     C.Y. Lum, K.-G. Lim, and K.-H. Chew, Revisiting the Kittel's model of antiferroelectricity: phase diagrams, hysteresis loops and electrocaloric effect J. Phys.: Condens. Matter **34,** 415702 (2022).

[22]     I. Zamaraite, R. Yevych, A. Dziaugys, A. Molnar, J. Banys, S. Svirskas, and Yu Vysochanskii. Double hysteresis loops in proper uniaxial ferroelectrics. Phys. Rev. Applied **10**, 034017 (2018).

[23]     I Suzuki, K Okada, Phenomenological theory of antiferroelectric transition. IV. Ferrielectric. J. Phys. Soc. Jap. **45(4),** 1302-1308 (1978).

[24]     L. Baudry, I. Lukyanchuk, V. M. Vinokur. Ferroelectric symmetry-protected multibit memory cell. Scientific Reports, **7,** Article number 42196 (2017). https://www.nature.com/articles/srep42196.pdf

[25]     Frederick Seitz, Henry Ehrenreich, and David Turnbull. Solid State Physics. (Academic Press 1996). pp. 80–150.

[26]     See supplementary Materials for details [URL will be provided by Publisher]

[27]     L. D. Landau, and I. M. Khalatnikov. On the anomalous absorption of sound near a second order phase transition point. In Dokl. Akad. Nauk SSSR **96**, 496 (1954).

[28]     Yu. M. Vysochanskii, M.M. Mayor, V. M. Rizak, V. Yu. Slivka, and M. M. Khoma. The tricritical Lifshitz point on the phase diagram of $Sn_2P_2(Se_xS_{1-x})_6$. Soviet Journal of Experimental and Theoretical Physics **95**, 1355 (1989).

[29]     A. Kohutych, R. Yevych, S. Perechinskii, V. Samulionis, J. Banys, and Yu. Vysochanskii. Sound behavior near the Lifshitz point in proper ferroelectrics. Phys. Rev. B **82**, 054101 (2010).

[30]     A. K. Tagantsev and G. Gerra. Interface-induced phenomena in polarization response of ferroelectric thin films. J. Appl. Phys. **100**, 051607 (2006).

[31]     P. Guranich, V.Shusta, E.Gerzanich , A.Slivka, I.Kuritsa, O.Gomonnai. "Influence of hydrostatic pressure on the dielectric properties of $CuInP_2S_6$ and $CuInP_2Se_6$ layered crystals." Journal of Physics: Conference Series **79**, 012009 (2007).

[32]     A. V. Shusta, A. G. Slivka, V. M. Kedylich, P. P. Guranich, V. S. Shusta, E. I. Gerzanich, I. P. Prits, Effect of uniaxial pressure on dielectric properties of $CuInP_2S_6$ crystals. Scientific Bulletin of Uzhhorod University. Physical series, **28**, 44 (2010).

[33]     A. Kohutych, V. Liubachko, V. Hryts, Yu. Shiposh, M. Kundria, M. Medulych, K. Glukhov, R. Yevych, and Yu. Vysochanskii. Phonon spectra and phase transitions in van der Waals ferroics MM'$P_2X_6$, Molecular Crystals and Liquid Crystals (2022). https://doi.org/10.1080/15421406.2022.2066787

[34]     J. Banys, J. Macutkevic, V. Samulionis, A. Brilingas & Yu. Vysochanskii, Dielectric and ultrasonic investigation of phase transition in $CuInP_2S_6$ crystals. Phase Transitions: A Multinational Journal **77:4**, 345 (2004).

[35]     V. Samulionis, J. Banys, Yu. Vysochanskii, and V. Cajipe. Elastic and electromechanical properties of new ferroelectric-semiconductor materials of $Sn_2P_2S_6$ family. Ferroelectrics **257:1**, 113 (2001).

[36]     A. N. Morozovska, M. D. Glinchuk, E.A. Eliseev. Phase transitions induced by confinement of ferroic nanoparticles. Phys. Rev. B **76**, 014102 (2007).

[37]     V. A. Shchukin and Dieter Bimberg. Reviews of Modern Physics, Vol. **71**, 1125 (1999).

[38]     Wenhui Ma. Appl. Phys. A **96**, 915-920 (2009).

[39] V. Ya. Shur, E. L. Rumyantsev, D. V. Pelegov, V. L. Kozhevnikov, E. V. Nikolaeva, E. L. Shishkin, A. P. Chernykh, and R. K. Ivanov. Barkhausen jumps during domain wall motion in ferroelectrics. Ferroelectrics, **267**, 347 (2002).

[40] E. A. Eliseev, Y. M. Fomichov, S. V. Kalinin, Yu. M. Vysochanskii, P. Maksymovich and A. N. Morozovska. Labyrinthine domains in ferroelectric nanoparticles: Manifestation of a gradient-induced morphological phase transition. Phys. Rev. B **98**, 054101 (2018).

[41] E. A. Eliseev, A. V. Semchenko, Y. M. Fomichov, M. D. Glinchuk, V. V. Sidsky, V. V. Kolos, Yu. M. Pleskachevsky, M. V. Silibin, N. V. Morozovsky, A. N. Morozovska. Surface and finite size effects impact on the phase diagrams, polar and dielectric properties of $(Sr,Bi)Ta_2O_9$ ferroelectric nanoparticles. J. Appl. Phys. **119**, 204104 (2016).

[42] A. K. Tagantsev and G. Gerra. Interface-induced phenomena in polarization response of ferroelectric thin films. J. Appl. Phys. **100**, 051607 (2006).

[43] L. D. Landau, E. M. Lifshitz, L. P. Pitaevskii. Electrodynamics of Continuous Media, (Second Edition, Butterworth-Heinemann, Oxford, 1984).

[44] A. N. Morozovska, E. A. Eliseev, A. Biswas, N. V. Morozovsky, and S. V. Kalinin. Effect of surface ionic screening on polarization reversal and phase diagrams in thin antiferroelectric films for information and energy storage. Phys. Rev. Applied **16**, 044053 (2021).

## Supplementary Materials
### Appendix A. Landau-Ginzburg-Devonshire approach

The density of the four-well LGD potential, $g_{LGD}$, which includes the Landau-Devonshire expansion in even powers of the polarization $P_3$ up to the eighth power, $g_{LD}$, the Ginzburg gradient energy $g_G$, and the elastic and electrostriction energies, $g_{ES}$, has the form [1, 2]:

$$g_{LGD} = g_{LD} + g_G + g_{ES}, \tag{S.1a}$$

$$g_{LD} = \frac{\alpha}{2}P_3^2 + \frac{\beta}{4}P_3^4 + \frac{\gamma}{6}P_3^6 + \frac{\delta}{8}P_3^8 - P_3 E_3, \tag{S.1b}$$

$$g_G = g_{33kl}\frac{\partial P_3}{\partial x_k}\frac{\partial P_3}{\partial x_l}, \tag{S.1c}$$

$$g_{ES} = -\frac{s_{ij}}{2}\sigma_i\sigma_j - Q_{i3}\sigma_i P_3^2 - Z_{i33}\sigma_i P_3^4 - W_{ij3}\sigma_i\sigma_j P_3^2. \tag{S.1d}$$

In accordance with LGD theory, the coefficient $\alpha$ depends linearly on the temperature $T$, $\alpha(T) = \alpha_T(T - T_C)$, where $T_C$ is the Curie temperature of the bulk ferrielectric. The coefficients $\beta$, $\gamma$, and $\delta$ in Eq.(S.1b) are temperature independent. The values $g_{33kl}$ are polarization gradient coefficients in the matrix notation, and the subscripts $k, l = 1 - 3$. The values $\sigma_i$ denote diagonal components of a stress tensor in the Voigt notation, and a subscript $i, j = 1 - 6$. The values $Q_{i3}$, $Z_{i33}$, and $W_{ij3}$ denote the components of a single linear and two nonlinear electrostriction strain tensors in the Voigt notation, respectively [3, 4]. $E_3$ is an electric field component co-directed with the polarization $P_3$, which is a superposition of external and depolarization fields.

Allowing for Khalatnikov relaxation, the minimization of Eq.(S.1a), $\frac{\partial g_{LGD}}{\partial P_3} = 0$, yields:

$$\Gamma\frac{\partial P_3}{\partial t} + [\alpha - 2\sigma_i(Q_{i3} + W_{ij3}\sigma_j)]P_3 + (\beta - 4Z_{i33}\sigma_i)P_3^3 + \gamma P_3^5 + \delta P_3^7 - g_{33kl}\frac{\partial^2 P_3}{\partial x_k \partial x_l} = E_3.$$
$$\tag{S.2a}$$

Here $\Gamma$ is the Khalatnikov kinetic coefficient [5], and the corresponding Landau-Khalatnikov relaxation time $\tau$ can be introduced as $\tau = \Gamma/|\alpha|$. The boundary condition for $P_3$ at the nanoparticle surface S is "natural", i.e., $g_{33lk}n_k\frac{\partial P_3}{\partial x_l}\Big|_S = 0$, where $\boldsymbol{n}$ is the outer normal to the surface.

The electric field $E_3 = -\frac{\partial \phi}{\partial x_3}$, and potential $\phi$ satisfies the Poisson equation inside the particle core,

$$\varepsilon_0\varepsilon_b\Delta\phi = \frac{\partial P_3}{\partial x_3}, \tag{S.2b}$$

and the Laplace equation outside the particle shell, $\Delta\phi = 0$. The 3D Laplace operator is denoted by the symbol $\Delta$. Equations (S.2b) are supplemented by the condition of potential continuity at the nanoparticle surface, $(\phi_{ext} - \phi_{int})|_S = 0$. The boundary condition for the normal components of electric displacements $\vec{D}$ is $\vec{n}(\vec{D}_{ext} - \vec{D}_{int})|_S = \rho_s$, where the surface charge density $\rho_s$ is given by the expression $\rho_s = -\varepsilon_0 \phi/\lambda$, and $\lambda$ is an effective screening length.

**A.1. Finite element modeling.** We perform finite element modeling (**FEM**) in COMSOL@MultiPhysics software, using electrostatics, solid mechanics, and general math (PDE toolbox) modules, for different discretization densities of the self-adaptive tetragonal mesh and polarization relaxation conditions. The size of the computational region is not less than 40×40×44 nm$^3$. The minimal size of a tetrahedral element in a mesh with fine discretization is equal to the minimal lattice constant, 0.65 nm, and the maximal size is 4 nm. The dependence on the mesh size is verified by increasing the minimal size from 0.6 nm to 2 nm. We verified that this results in minor changes in the electric polarization, electric field, and elastic stress and strain, such that the spatial distribution of each of these quantities becomes less smooth (i.e., they contain numerical errors in the form of a small random noise).

**A.2. Single-domain approximation.** The strength and anisotropy of the polarization gradient energy are defined by the tensor $g_{33ij}$. As it has been shown earlier [6], the critical sizes of the CIPS core corresponding to the ferroelectric-paraelectric phase transition, as well as the domain structure appearance and its morphology in the ferroelectric state, depend strongly on the magnitude and anisotropy of the polarization gradient coefficients, $g_{33ij}$. In particular, several times a small increase in $g_{33ij}$ (e.g., above $10^{-9}$ J m$^3$/C$^2$) leads to a relatively small increase of the critical sizes (less than several nm); but it strongly suppresses the domain structure appearance in the ferrielectric state. For instance, in the case $g \cong 2 \times 10^{-9}$ J m$^3$/C$^2$ (used in this work, see **Table I**) CIPS nanoparticles are mostly single-domain above the critical sizes and paraelectric below the critical sizes. In addition, for the case of natural boundary conditions, $g_{33kl} n_k \frac{\partial P_3}{\partial x_l} = 0$, also used in this work, polarization gradient effects can be neglected in the single-domain state.

The field dependence of a quasi-static single-domain polarization can be found from the following equation:

$$\Gamma \frac{\partial P_3}{\partial t} + \alpha^* P_3 + \beta^* P_3^3 + \gamma P_3^5 + \delta P_3^7 = E. \quad (S.3)$$

Here $\beta^*(\sigma_i) = \beta - 4Z_{i33}\sigma_i$, and $E$ is an external field inside the core, whose frequency $\omega$ is regarded to be very small, e.g., $\omega\tau \ll 10^{-4}$.

The depolarization field $E_d$ and stresses $\sigma_i$ contribute to the "renormalization" of coefficient $\alpha(T)$, which becomes the temperature-, stress-, shape-, size-, and screening-dependent function $\alpha^*$ [1]:

$$\alpha^*(T, n_d, \sigma_i) = \alpha(T) + \frac{n_d}{\varepsilon_0[\varepsilon_b n_d + \varepsilon_s(1-n_d) + n_d(L/\lambda)]} - 2\sigma_i(Q_{i3} + W_{ij3}\sigma_j). \quad (S.4)$$

The derivation of the second term in Eq.(S.4) is given in Ref.[7]. It is based on the approximate expression for electric field $E_3$ inside the ellipsoidal particle, $E_3 = \frac{\varepsilon_0 \varepsilon_s E_0 - n_d P_3}{\varepsilon_0[\varepsilon_b n_d + \varepsilon_s(1-n_d) + n_d(L/\lambda(E_3))]}$, where $E_0$ is a homogeneous external field far from the particle. Parameters $\varepsilon_b$ and $\varepsilon_s$ are the background dielectric permittivity [8] of the ferrielectric core and the relative dielectric permittivity of its screening shell or surrounding, respectively. The dimensionless parameter $n_d$ is the shape-dependent depolarization factor, introduced as [9]:

$$n_d(\zeta) = \frac{1-\xi^2}{\xi^3}\left(\ln\sqrt{\frac{1+\xi}{1-\xi}} - \xi\right), \quad \xi(\zeta) = \sqrt{1-\zeta^2}, \quad \zeta = \frac{R}{L}. \tag{S.5}$$

Here $\xi$ is the eccentricity ratio of the ellipsoid with semi-axes $R$ and $L$, and $\zeta$ is the dimensionless shape factor.

Note that the linear dependence of the surface charge density $\rho_s$ on the local potential is a simplification required for analytical calculations. It is well known that a surface state model predicts a nonlinear dependence of the surface charge on the surface potential [10]. In what follows we approximately account for the fact that $\lambda$ is strongly field-dependent in ferrielectric-semiconductors, such as CIPS; it is the consequence of field effects in semiconductors, which is caused by the band bending near the surface. The band bending, and so the field effect, is small for a small homogeneous external field $E_0$, and therefore the concentration of free charges, induced by $E_0$, is relatively small inside the core. At the same time a "bare" (i.e., unscreened) depolarization field, $E_{d0} = \frac{-n_d P_3}{\varepsilon_0[\varepsilon_b n_d + \varepsilon_s(1-n_d)]}$, significantly exceeds the thermodynamic coercive field in magnitude near the core surface. The strong band bending occurs in response to the field $E_{d0}$, and the field-induced concentration of carriers becomes very high at the surface. Since the concentration determines the effective screening length $\lambda$, we must take into account that $\lambda$ is significantly different for the external and depolarization fields, namely $\lambda(E_0) \gg \lambda(E_d)$. Hence the "screened" depolarization field, $E_d = \frac{-n_d P_3}{\varepsilon_0[\varepsilon_b n_d + \varepsilon_s(1-n_d) + n_d(L/\lambda(E_d))]}$, where $\lambda(E_d)$ is rather small, becomes significantly smaller than the bare field $E_{d0}$. At the same time the external field, $E = \frac{\varepsilon_s E_0}{\varepsilon_b n_d + \varepsilon_s(1-n_d) + n_d(L/\lambda(E_0))}$, remains almost unscreened, namely $E \approx \frac{\varepsilon_s E_0}{\varepsilon_b n_d + \varepsilon_s(1-n_d)} \cong E_0$, if we regard that $\lambda(E_0) \gg L$ and $\varepsilon_b = \varepsilon_s$ (as suggested in **Table S1**). Also, we assume that the screening charge distribution adiabatically follows polarization changes (i.e., $\vec{E}_d$ changes), and is much less sensitive to the $\vec{E}_0$ changes.

The values of $T_C$, $\alpha_T$, β, γ, δ, $Q_{i3}$, and $Z_{i33}$ have been determined in Refs.[11] from the fitting of temperature dependent experimental data for the dielectric permittivity [12, 13, 14], spontaneous polarization [15], and lattice constants [16] as a function of hydrostatic pressure.

Elastic compliances, $s_{ij}$, were estimated from ultrasound velocity measurements [17, 18]. The details for determining the CIPS material parameters are given in the Supplement of Ref.[1]. Using the experimental results of Ref.[19], we managed [2] to estimate the diagonal components, $W_{ii3}$, which are coupled with the hydrostatic pressure in Eq.(S.1d). The gradient coefficients, $g_{33ij}$, are not determined experimentally, but they are estimated from the width of the domain walls. The CIPS parameters are listed in **Table SI.**

**Table SI.** LGD parameters for a bulk ferrielectric CuInP$_2$S$_6$.

| coefficient | value |
| --- | --- |
| $\varepsilon_b \approx \varepsilon_s$ | 9 |
| $\alpha_T$ (C$^{-2}$·m J/K) | 1.64067×10$^7$ |
| $T_C$ (K) | 292.67 |
| $\beta$ (C$^{-4}$·m$^5$J) | 3.148×10$^{12}$ |
| $\gamma$ (C$^{-6}$·m$^9$J) | $-$1.0776×10$^{16}$ |
| $\delta$ (C$^{-8}$·m$^{13}$J) | 7.6318×10$^{18}$ |
| $Q_{i3}$ (C$^{-2}$·m$^4$) | $Q_{13} = 1.70136 - 0.00363\,T$, $Q_{23} = 1.13424 - 0.00242\,T$, $Q_{33} = -5.622 + 0.0105\,T$ |
| $Z_{i33}$ (C$^{-4}$·m$^8$) | $Z_{133} = -2059.65 + 0.8\,T$, $Z_{233} = -1211.26 + 0.45\,T$, $Z_{333} = 1381.37 - 12\,T$ |
| $W_{ij3}$ (C$^{-2}$·m$^4$ Pa$^{-1}$) | $W_{113} \approx W_{223} \approx W_{333} \cong -2\times 10^{-12}$ |
| $s_{ij}$ (Pa$^{-1}$) | $s_{11} = 1.510\times 10^{-11}$, $s_{12} = 0.183\times 10^{-11}$ |
| $g_{33ij}$ (J m$^3$/C$^2$) | $g \cong 2\times 10^{-9}$ |

### Appendix B. Algorithm for loop classification

Numerical results presented in the work are obtained and visualized using a specialized software, Mathematica 13.1 [20]. The Mathematica notebook, which contain the codes, is available per reasonable request.

For constructing phase diagrams, we found the minima of the free energy (S.1) numerically as a function of several variables (e.g., temperature, pressure, and/or sizes), and calculated the equilibrium values of the polarization in the minima. Whenever it was possible, we used analytical expressions for phase boundaries (e.g., for the stability of the paraelectric phase), and superimposed corresponding curves on the color maps of polarization.

The inflection points of the quasi-static single-domain loop can be found from the static Eq.(S.3), which yields the following equation:

$$\alpha^* + 3\beta^* P_3^2 + 5\gamma P_3^4 + 6\delta P_3^6 = 0. \tag{S.6}$$

For constructing loop diagrams, six possible solutions of Eq.(S.6) for polarization, $\pm P_i$, where $i = 1,2,3$, are analyzed. The real and positive solutions, which maximal amount is three, can always be ordered as $0 < P_1 < P_2 < P_3$. Corresponding values of electric field, $E(P_1)$, $E(P_2)$ and $E(P_3)$, are determined from the static Eq.(S.3) and denoted as $E_1$, $E_2$, and $E_3$. Analyzing the amount of $P_i$ and the ordering of the corresponding $E_i$, our algorithm distinguishes the following types of polarization curves and loops.

### 1. The region of paraelectric curves (PC)

In the hysteresis-less region, real solutions are absent in Eq.(S.6), and all $P_i$ have an imaginary part:

$$Im(P_i) \neq 0 \quad \text{for all roots "i".} \quad (S.7)$$

Any type of hysteresis is absent in the PC region, and a typical polarization field dependence is shown in **Fig. S1.**

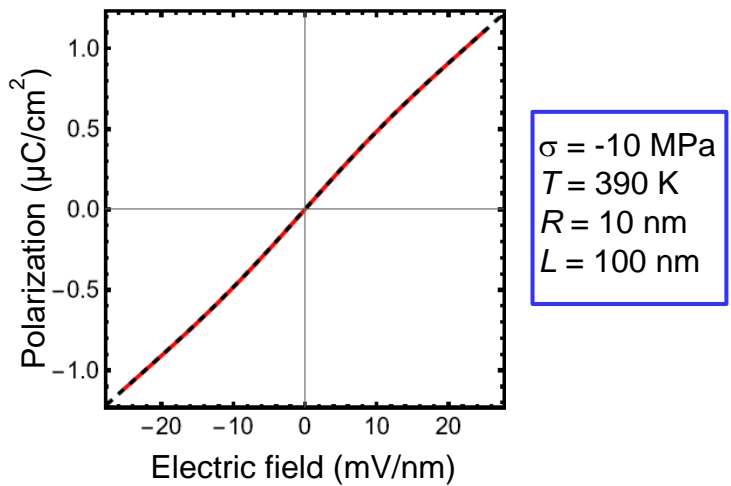

**Figure S1.** Electric field dependence of the polarization $P_3$, calculated for CIPS nanoellipsoids with sizes, temperature, and hydrostatic pressure listed in the legend, and a screening length $\lambda = 0.5$ nm. Black dashed and red solid curves, which coincide, are static and quasi-static dependences calculated for $\omega\tau = 10^{-5}$, respectively.

### 2. The region of double loops (DL)

In the case of DL, which are shown in **Fig. S2**, we have four roots with zero imaginary parts in Eq.(S.6), which are $\pm P_1$ and $\pm P_2$.

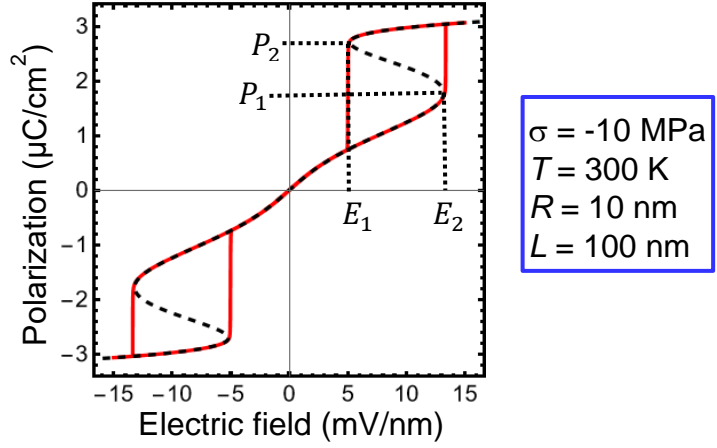

**Figure S2.** Electric field dependence of the polarization $P_3$, calculated for CIPS nanoellipsoids with sizes, temperature, and hydrostatic pressure listed in the legend, and a screening length $\lambda = 0.5$ nm. The black dashed curve is a static dependence, the quasi-static red solid loop is calculated for $\omega\tau = 10^{-5}$.

The following conditions of DL appearance can be derived from **Fig. S2**:

$$0 < P_1 < P_2, \quad E_1 > 0, \text{ and } E_2 > 0. \tag{S.8}$$

### 3. The region of triple loops (TL)

In the most interesting case of TL, shown in **Fig. S3**, all six solutions in Eq.(S.6), $\pm P_1$, $\pm P_2$ and $\pm P_3$, are real. The condition for the roots is the following:

$$0 < P_1 < P_2 < P_3, \quad E_1 < 0, E_2 > 0, \text{ and } E_3 > 0. \tag{S.9}$$

Corresponding static curves have a complex structure with three well-separated unstable regions.

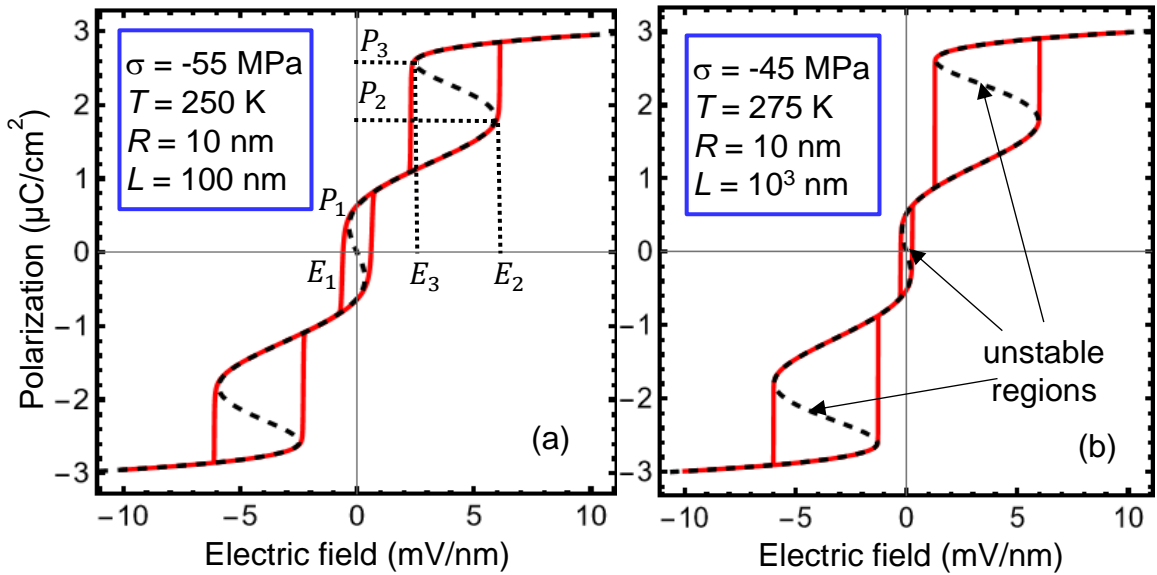
6

**Figure S3.** Electric field dependence of the polarization $P_3$, calculated for CIPS nanoellipsoids with two different sizes, temperatures, and hydrostatic pressures listed in the legends of plots **(a)** and **(b)**, and a screening length $\lambda = 0.5$ nm. Black dashed curves are static dependences, red solid loops are quasi-static hysteresis loops calculated for $\omega\tau = 10^{-5}$. Plot **(a)** shows the onset of TL, and plot **(b)** shows its disappearance.

### 4. The region of pinched single loops of the first type (PL-I)

The case of PL-I is shown by a red color in **Fig. S4**, and the static curve is shown by a black dashed line in **Fig. S4**. This feature has a complex structure with two unstable regions and one metastable central region. In this case, there are four real roots in Eq.(S.6), which are $\pm P_1$ and $\pm P_2$. The roots with lower polarization values correspond to the higher values of the critical field $E_i$. They are situated in such a way, that the loop is pinched in the middle (see **Fig. S4**). From the figure, we get the following set of conditions for $E_i$:

$$0 < P_1 < P_2, \quad E_1 > 0, \text{ and } E_2 < 0. \tag{S.10}$$

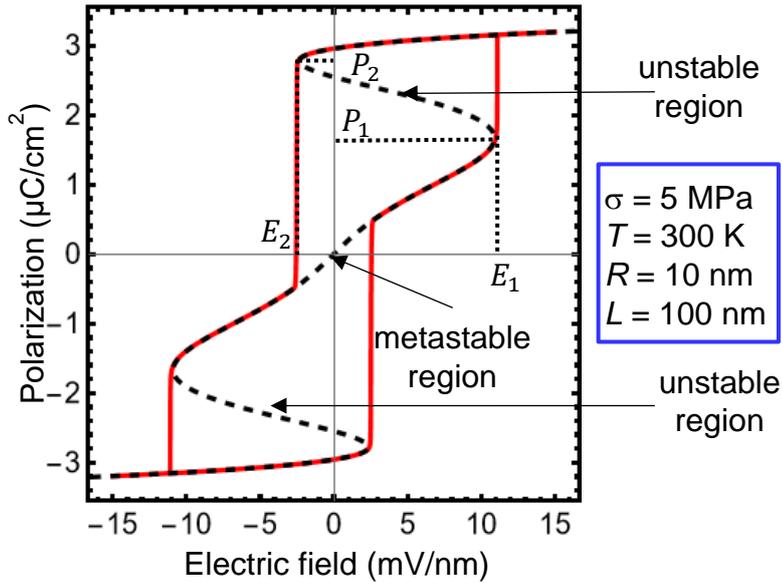

**Figure S4.** Electric field dependence of the polarization $P_3$, calculated for CIPS nanoellipsoids with sizes, temperature, and hydrostatic pressure listed in the legend, and a screening length $\lambda = 0.5$ nm. The black dashed curve is a static dependence, the quasi-static red solid loop is calculated for $\omega\tau = 10^{-5}$.

### 5. The region of pinched loops of the second (PL-II) and third (PL-III) types

Two possible shapes of pinched loops, PL-II and PL-III, are shown in **Fig. S5(a)** and **S5(b)**, respectively. Corresponding static curves, shown by black dashed lines in **Fig. S5(a)** and **S5(b)**, have a complex structure with three unstable regions. In both cases of PL-II and PL-III, there are six real roots in Eq.(S.6), which are $\pm P_1$, $\pm P_2$ and $\pm P_3$. The case of PL-II, shown in **Fig. S5(a)**, has the following appearance conditions:

$$0 < P_1 < P_2 < P_3, \quad E_1 < 0, E_2 > 0, E_3 < 0, \text{ and } E_2 > -E_3. \tag{S.11a}$$

The case of PL-III, shown in **Fig. S5(b)**, has the following appearance conditions:

$$0 < P_1 < P_2 < P_3, \quad E_1 < 0, \text{ and } -E_1 > |E_3|. \tag{S.11b}$$

The conditions (S.11) reflect the principal difference between the structure of static curves in **Fig. S5(a)** and **S5(b)**, respectively.

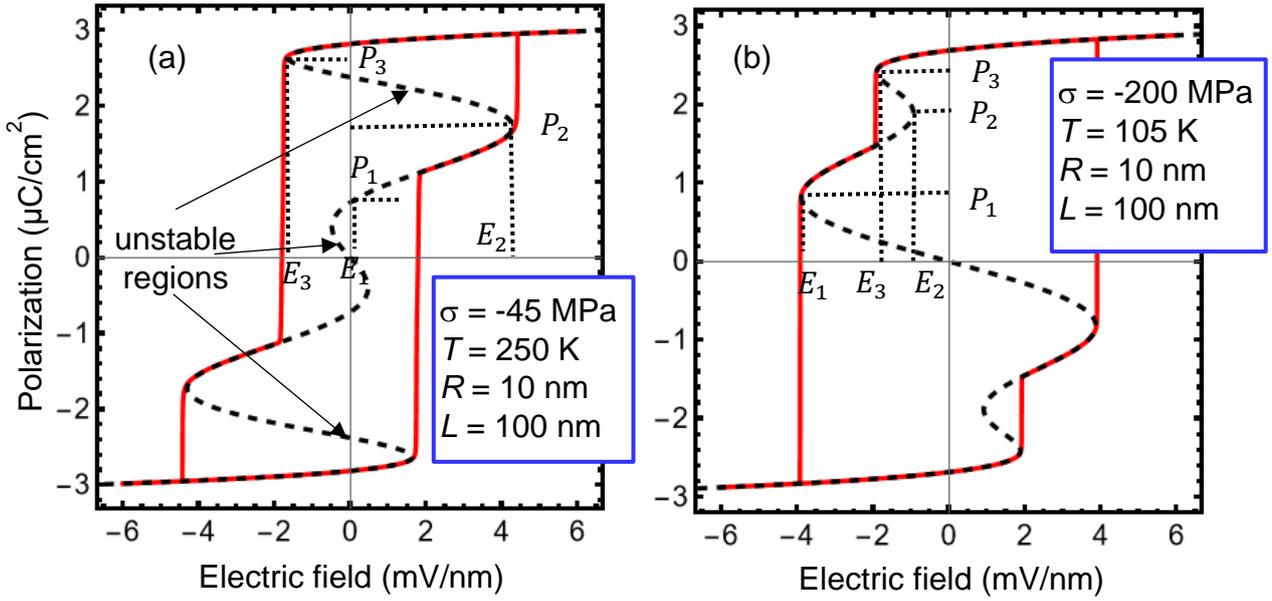

**Figure S5.** Electric field dependence of the polarization $P_3$, calculated for CIPS nanoellipsoids with sizes, temperatures, and hydrostatic pressures listed in the legends of plots **(a)** and **(b)**, and a screening length $\lambda = 0.5$ nm. Black dashed curves are static dependences, red solid loops are quasi-static hysteresis loops calculated for $\omega\tau = 10^{-5}$. Plot **(a)** shows the PL-II, and plot **(b)** shows the PL-III.

### 6. The region of single loops of the first type (SL-I)

In the case of SL-I we have either six, or four real roots in Eq.(S.6). Corresponding static curves, shown by black dashed lines in **Fig. S6**, have a complex structure with either three, or two unstable regions. Compared to the previous cases, the lower roots $P_i$ correspond to the lower critical field (see **Fig. S6**). The following conditions should be valid for the cases shown in **Fig. S6(a)** and **S6(b)**:

$$0 < P_1 < P_2 < P_3, \quad E_1 > 0, \quad E_3 < 0, \quad E_1 > E_3, \text{ and } |E_2| < -E_3. \tag{S.12a}$$

The following conditions should be valid for the case shown in **Fig. S6(c)**:

$$0 < P_1 < P_2, \quad E_1 > 0, \quad E_2 < 0, \text{ and } -E_2 > E_1. \tag{S.12b}$$

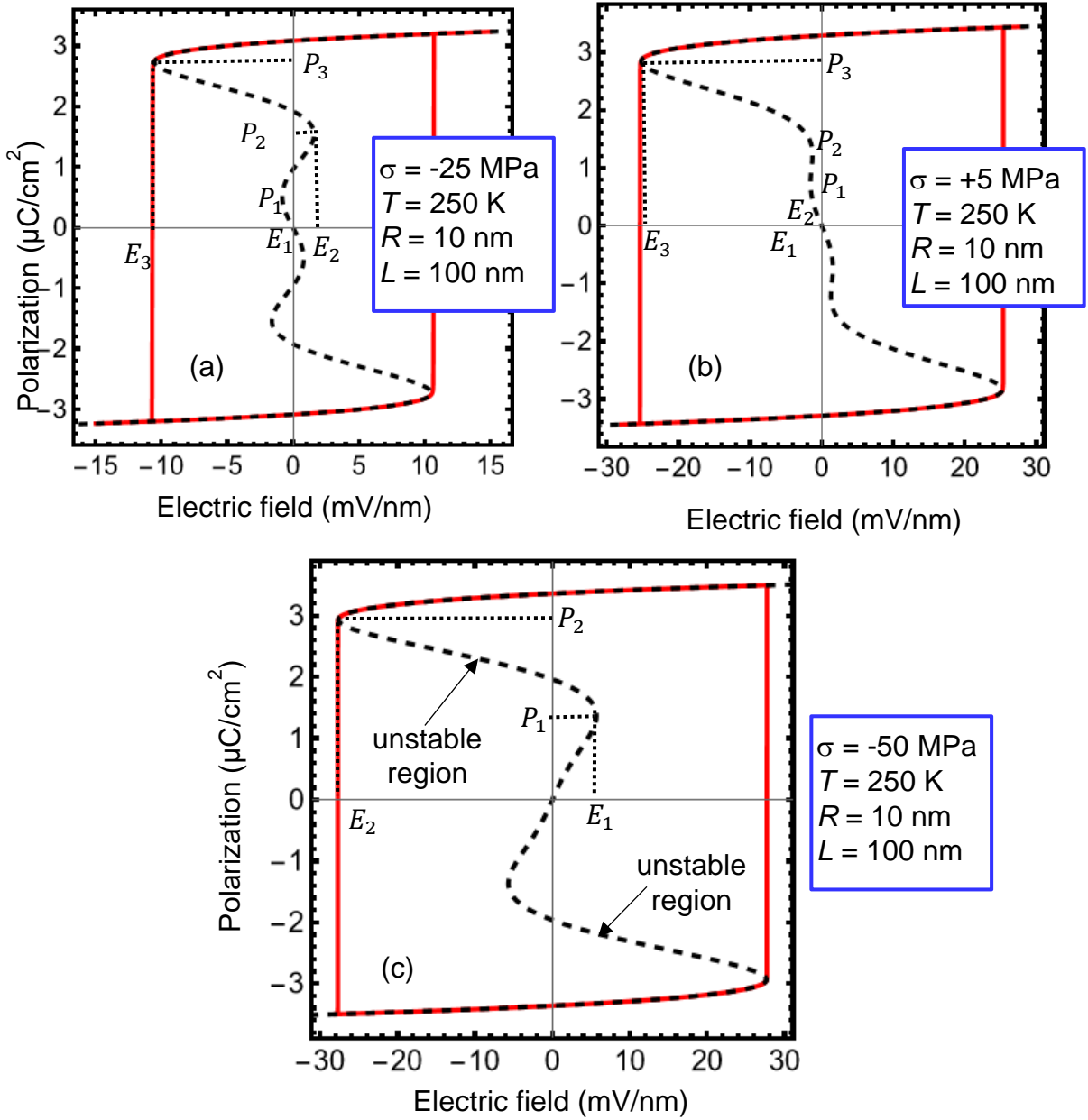

**Figure S6.** Electric field dependence of the polarization $P_3$, calculated for CIPS nanoellipsoids with sizes, temperatures, and hydrostatic pressure listed in the legends of plots **(a)**, **(b)**, and **(c)**, and a screening length $\lambda = 0.5$ nm. Black dashed curves are static dependences, red solid loops are quasi-static hysteresis loops calculated for $\omega\tau = 10^{-5}$. Plots **(a)** and **(b)** show the case when Eq.(S.6) has six real roots, and plot **(c)** shows the case when Eq.(S.6) has four real roots.

### 7. The region of single loops of the second type (SL-II)

The case of SL-II is shown by a red color in **Fig. S7**, and the static curve is shown by a black dashed line in **Fig. S7**. This feature has a simple structure with one unstable region in the center. In this case, there are only two real roots in Eq.(S.6), which are $\pm P_1$. In our algorithm, we can identify SL by the number of roots, since it is the only region with two real roots.

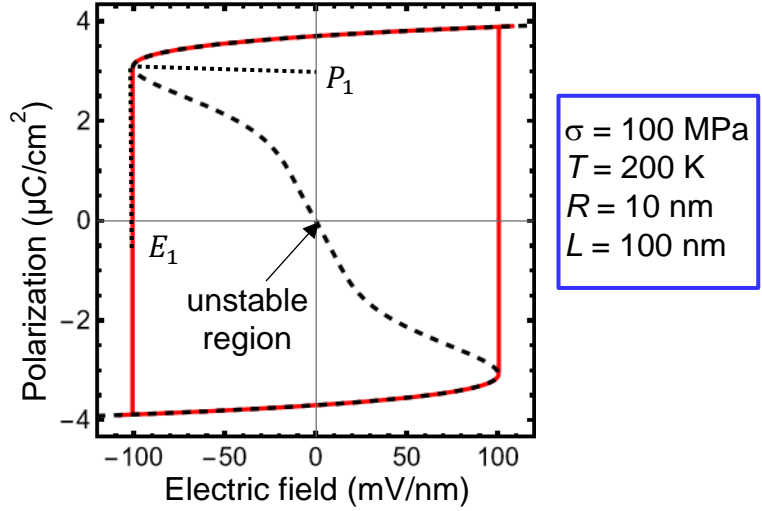

**Figure S7.** Electric field dependence of the polarization $P_3$, calculated for CIPS nanoellipsoids with sizes, temperature, and hydrostatic pressure listed in the legend, and a screening length $\lambda = 0.5$ nm. The black dashed curve is a static dependence, the quasi-static red solid loop is calculated for $\omega\tau = 10^{-5}$.

Another important thing we need to determine are the lines separating the regions (phases) in the pressure-temperature diagrams, shown in **Figs. 3-5** in the main part of the paper. Note, that the transitions between different phases occur when the number of the real roots in Eq.(S.6) changes. In particular, two roots "merge" with an inflection point, as shown schematically in **Fig. S8.**

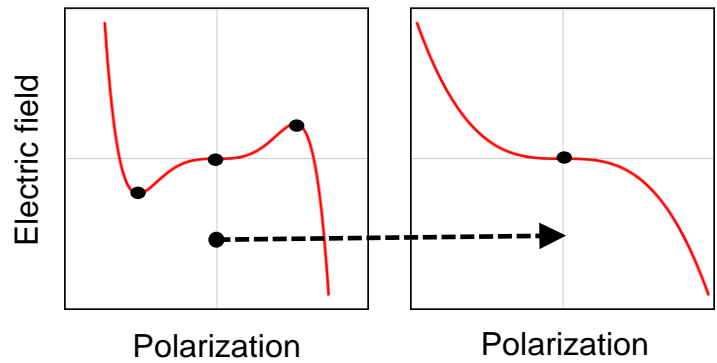

**Figure S8**. The transition from two extrema and one inflection point to the inflection point.

This transformation can be analytically described as the graphical solution of two coupled equations, one is Eq.(S.6) – shown in Eq.(S.13a), and another equation is the zero condition of the second derivative of (S.3) – shown in Eq.(S.13b):

$$\alpha^* + 3\beta^* P_3^2 + 5\gamma P_3^4 + 6\delta P_3^6 = 0, \qquad (S.13a)$$

$$6\beta^* P_3 + 20\gamma P_3^3 + 36\,\delta P_3^5 = 0. \qquad (S.13b)$$

The graphical solutions of these two equations yield the dashed and dashed-dotted curves in the phase diagrams in **Figs. 3-5**, shown in the main part of the paper. The dotted curves in the phase diagrams are given by equation $\alpha^* = 0$.

**Appendix C. The pressure influence on polarization switching at room temperature**

Here we analyze the pressure influence on polarization switching at room temperature, since the temperature is required for most applications. For CIPS nanodisks, the pressure-induced transition from double to single hysteresis loops occurs at relatively high compressive pressures (see **Fig. S9**). Under the increase of pressure from 70 MPa to 170 MPa, small double loops grow and become more pronounced, finally merging together into a pinched loop of the second type. Further increase of pressure from 170 MPa to 230 MPa leads to a gradual growth of the pinched section width, and the pinched loop eventually transforms into a single loop of the first type. Two coercive fields of double loops vary in the range (0 – 60) mV/nm and decrease with an increase in pressure. The coercive field of pinched and single loops increases with a pressure increase from 20 mV/nm to 60 mV/nm.

For CIPS nanospheres, the pressure-induced transition from double to pinched hysteresis loops occurs under the pressure increase from 20 MPa to 100 MPa (see **Fig. S10**). Further increase of pressure from 100 MPa to 140 MPa leads to a gradual growth of the pinched section, and the pinched loop eventually transforms into a single loop of the second type. The coercive fields are two times smaller than the fields for nanodisks. For CIPS nanoneedles, the pressure-induced transition from double hysteresis loops to pinched, and then to single loops, occurs within a narrower pressure range, from -40 MPa to 10 MPa (see **Fig. S11**). The coercive fields are approximately two times smaller than the fields for nanospheres.

**Figure S12** shows the onset and temperature evolution of stress-induced triple hysteresis loops in CIPS nanoneedles in the temperature range (265 – 300) K. Under zero surface tension ($\mu = 0$), the triple hysteresis loop appears from pinched loops of the third type in the temperature range (265-270) K for pressure $\sigma \cong -50$ MPa, and disappears near 280 K [see **Figs. S12(a)-(c)**]. A single hysteresis loop is stable for a stress-free nanoneedle, as shown in **Figs. S12(d)**. Negative surface tensions, $\mu \cong -(0.7 - 1)$N/m, induce triple hysteresis loops around (285-300) K for zero external pressure, $\sigma = 0$ [see **Figs. S12(e)-(j)**].

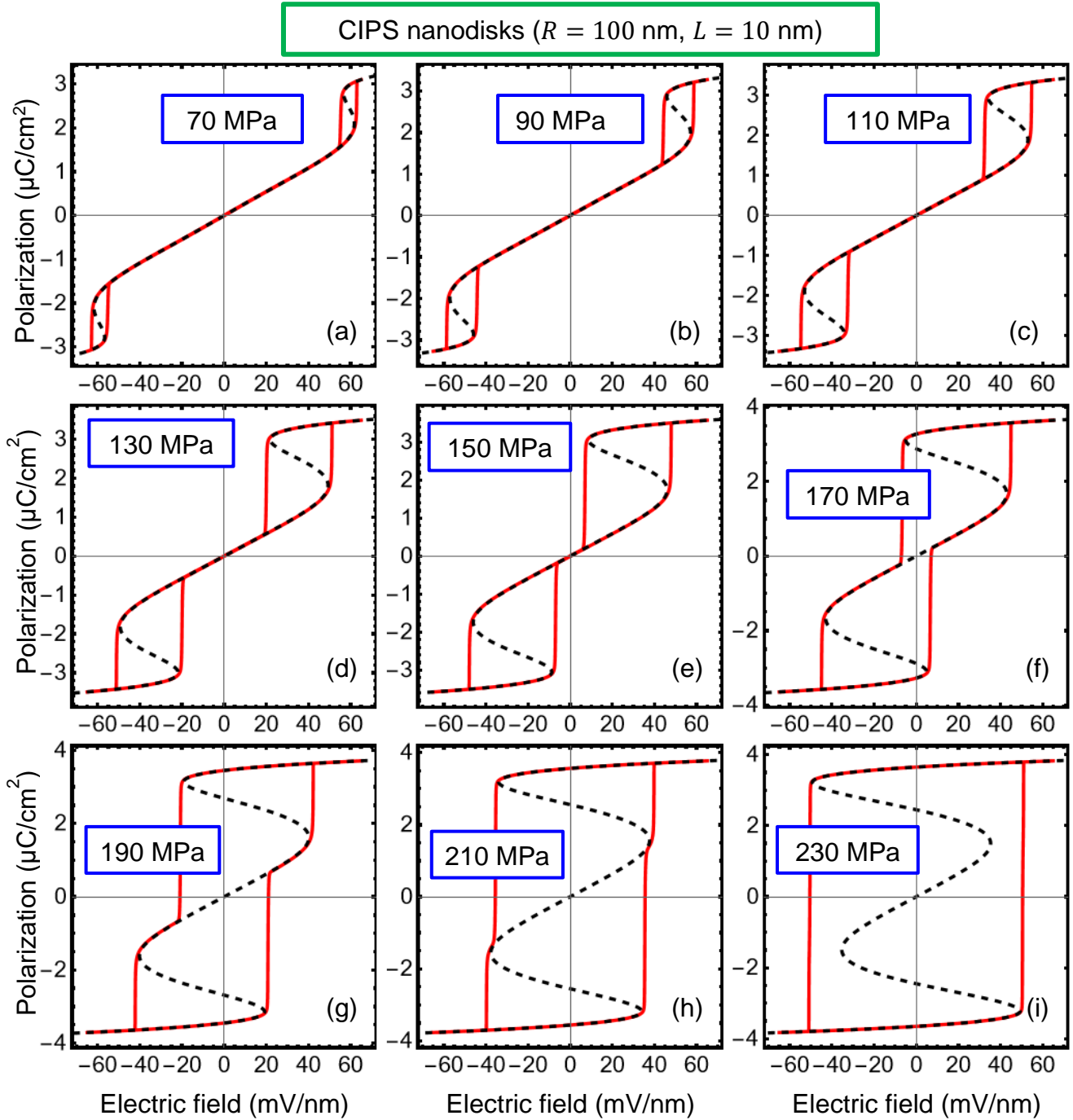

**FIGURE S9**. Electric field dependence of the polarization $P_3$, calculated for CIPS nanodisks with $R = 100$ nm, $L = 10$ nm, temperature $T = 300$ K, screening length $\lambda = 0.5$ nm, and different hydrostatic pressures: $\sigma = 70$ MPa **(a)**, 90 MPa **(b)**, 110 MPa **(c)**, 130 MPa **(d)**, 150 MPa **(e)**, 170 MPa **(f)**, 190 MPa **(g)**, 210 MPa **(h)**, and 230 MPa **(i)**. Black dashed curves are static dependences, red solid loops are quasi-static hysteresis loops calculated for $\omega\tau = 10^{-5}$. Other parameters are listed in **Table SI**.

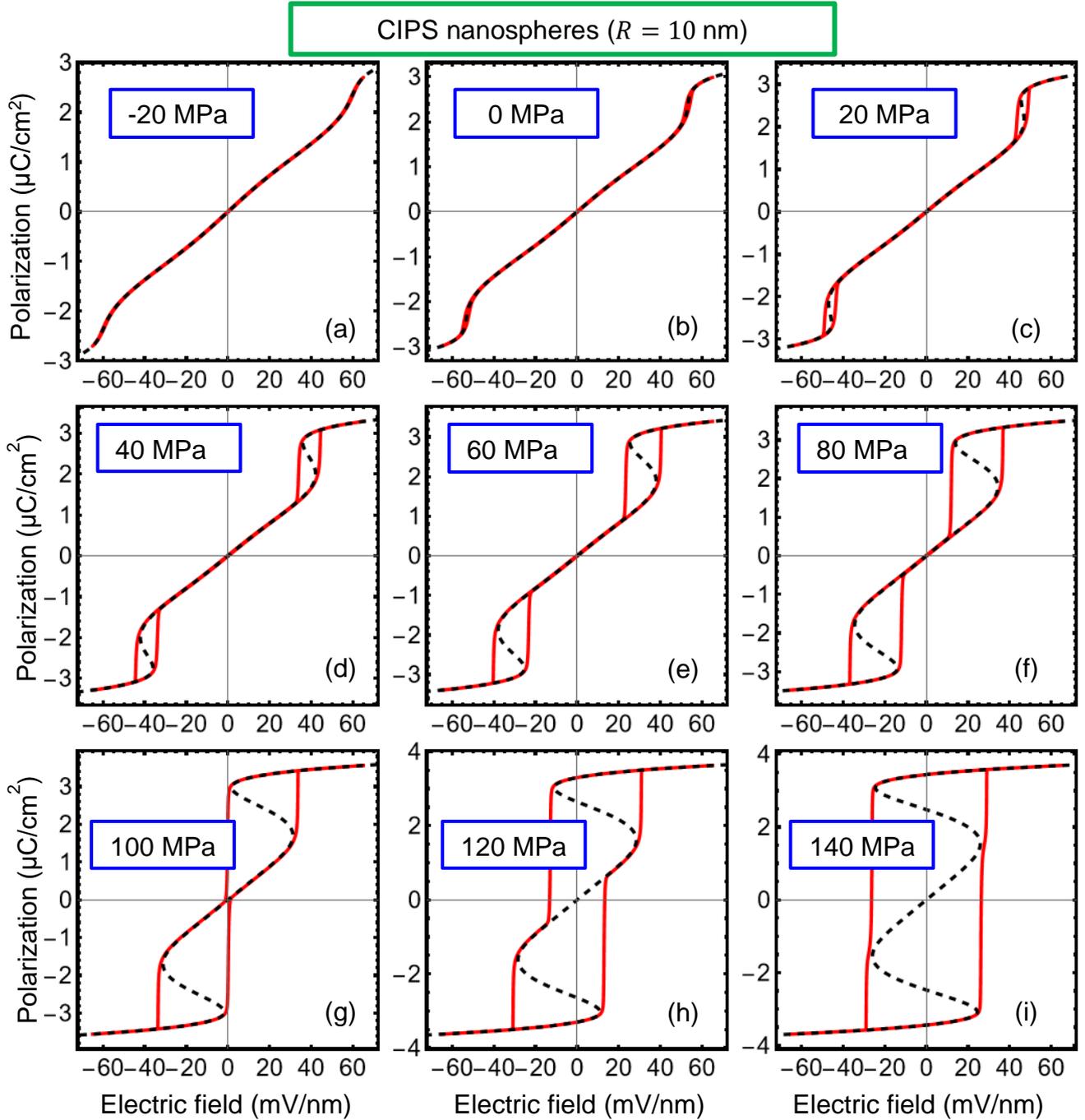

**FIGURE S10**. Electric field dependence of the polarization $P_3$, calculated for CIPS nanospheres with $R = 10$ nm, temperature $T = 300$ K, screening length $\lambda = 0.5$ nm, and different hydrostatic pressures: $\sigma = $ -20 MPa **(a)**, 0 **(b)**, 20 MPa **(c)**, 40 MPa **(d)**, 60 MPa **(e)**, 80 MPa **(f)**, 100 MPa **(g)**, 120 MPa **(h)**, and 140 MPa **(i)**. Black dashed curves are static dependences, red solid loops are quasi-static hysteresis loops calculated for $\omega\tau = 10^{-5}$. Other parameters are listed in **Table SI**.

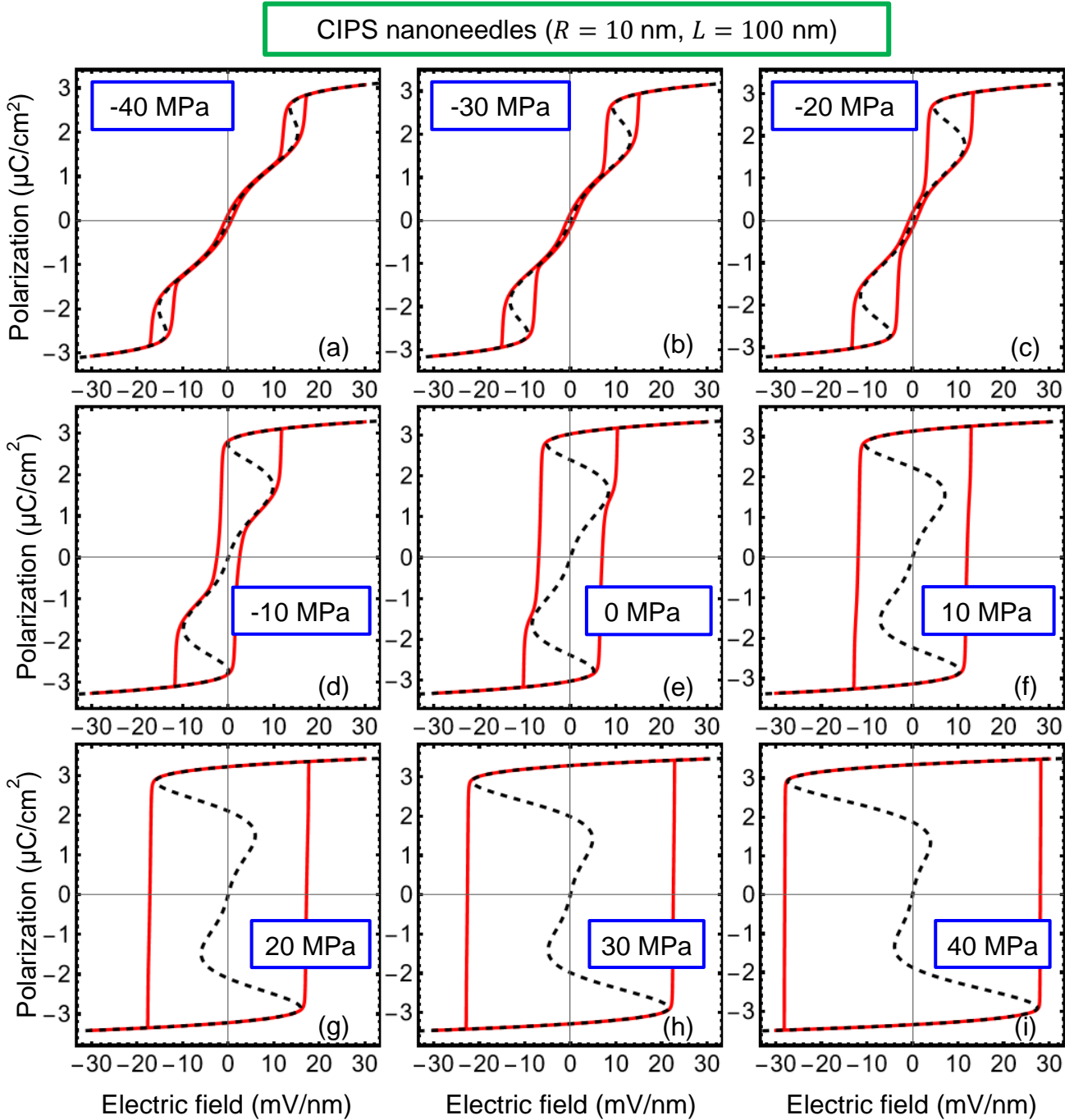

**FIGURE S11.** Electric field dependence of the polarization $P_3$, calculated for CIPS nanoneedles with $R = 10$ nm, $L = 100$ nm, temperature $T = 300$ K, screening length $\lambda = 0.5$ nm, and different hydrostatic pressures: $\sigma =$ -40 MPa **(a)**, -30 MPa **(b)**, -20 MPa **(c)**, -10 MPa **(d)**, 0 MPa **(e)**, 10 MPa **(f)**, 20 MPa **(g)**, 30 MPa **(h)**, and 40 MPa **(i)**. Black dashed curves are static dependences, red solid loops are quasi-static hysteresis loops calculated for $\omega\tau = 10^{-5}$. Other parameters are listed in **Table SI**.

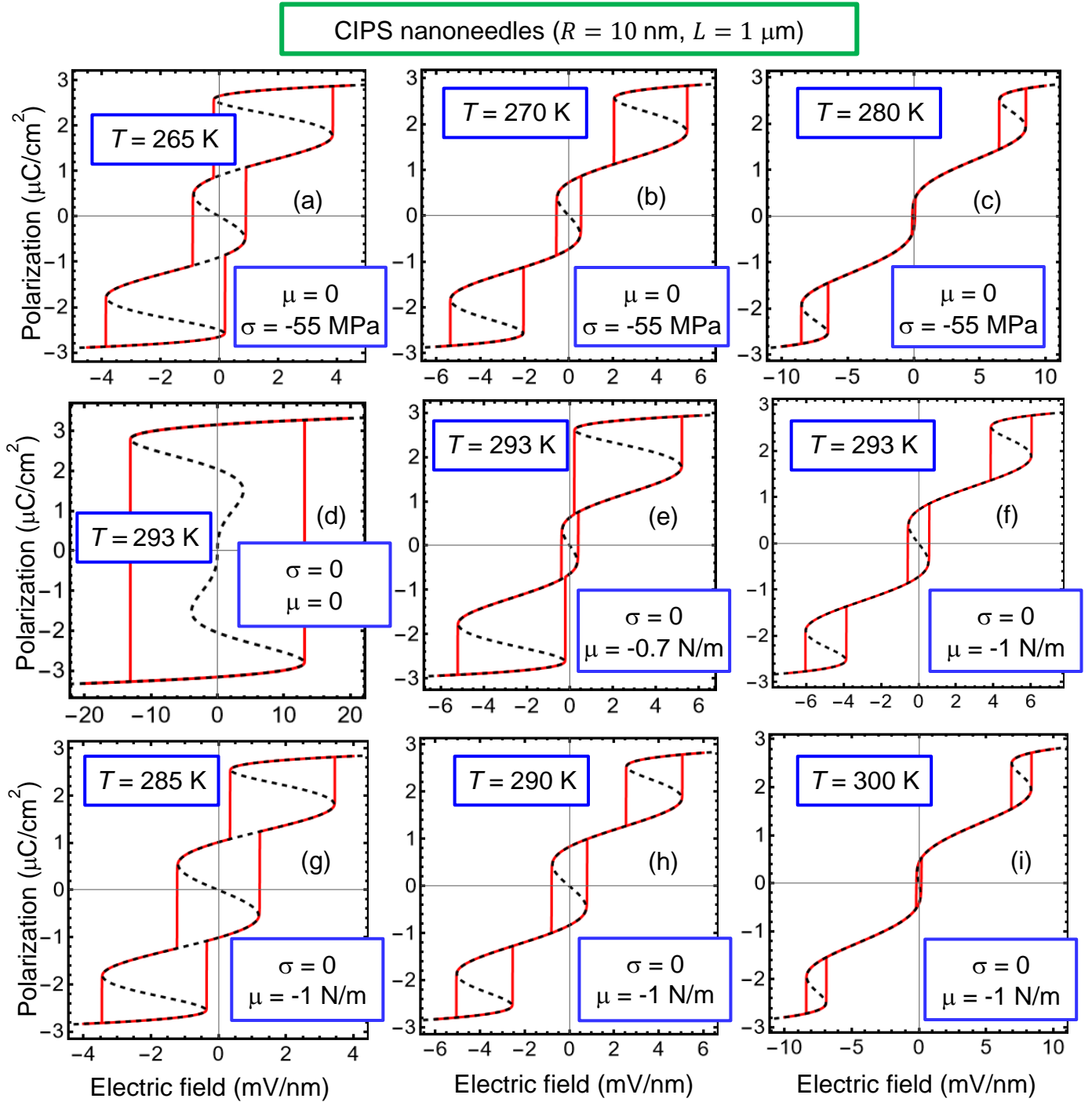

**FIGURE S12.** Electric field dependence of the polarization $P_3$, calculated for CIPS nanoneedles with sizes $R = 10$ nm and $L = 1$ μm. Temperature $T$, hydrostatic pressure $\sigma$, and surface tension coefficient $\mu$, are listed in the legends. Black dashed curves are static dependences, red solid loops are quasi-static hysteresis loops calculated for $\omega\tau = 10^{-5}$. The screening length $\lambda = 0.5$ nm, other parameters are listed in **Table SI**.

# References

[1] A. N. Morozovska, E. A. Eliseev, S. V. Kalinin, Y. M. Vysochanskii, and Petro Maksymovych. Stress-Induced Phase Transitions in Nanoscale $CuInP_2S_6$. Phys. Rev. B **104**, 054102 (2021).

[2] A. N. Morozovska, E. A. Eliseev, Yu. M. Vysochanskii, V. V. Khist, and D. R. Evans. Screening-Induced Phase Transitions in Core-Shell Ferroic Nanoparticles. Phys. Rev. Materials **6**, 124411 (2022).

[3] Yu. M. Vysochanskii, M.M. Mayor, V. M. Rizak, V. Yu. Slivka, and M. M. Khoma. The tricritical Lifshitz point on the phase diagram of $Sn_2P_2(Se_xS_{1-x})_6$. Soviet Journal of Experimental and Theoretical Physics **95**, 1355 (1989).

[4] A. Kohutych, R. Yevych, S. Perechinskii, V. Samulionis, J. Banys, and Yu. Vysochanskii. Sound behavior near the Lifshitz point in proper ferroelectrics. Phys. Rev. B **82**, 054101 (2010).

[5] L. D. Landau, and I. M. Khalatnikov. On the anomalous absorption of sound near a second order phase transition point. In Dokl. Akad. Nauk SSSR **96**, 496 (1954).

[6] E. A. Eliseev, Y. M. Fomichov, S. V. Kalinin, Yu. M. Vysochanskii, P. Maksymovich and A. N. Morozovska. Labyrinthine domains in ferroelectric nanoparticles: Manifestation of a gradient-induced morphological phase transition. Phys. Rev. B **98**, 054101 (2018).

[7] E. A. Eliseev, A. V. Semchenko, Y. M. Fomichov, M. D. Glinchuk, V. V. Sidsky, V. V. Kolos, Yu. M. Pleskachevsky, M. V. Silibin, N. V. Morozovsky, A. N. Morozovska. Surface and finite size effects impact on the phase diagrams, polar and dielectric properties of $(Sr,Bi)Ta_2O_9$ ferroelectric nanoparticles. J. Appl. Phys. **119**, 204104 (2016).

[8] A. K. Tagantsev and G. Gerra. Interface-induced phenomena in polarization response of ferroelectric thin films. J. Appl. Phys. **100**, 051607 (2006).

[9] L. D. Landau, E. M. Lifshitz, L. P. Pitaevskii. Electrodynamics of Continuous Media, (Second Edition, Butterworth-Heinemann, Oxford, 1984).

[10] Frederick Seitz, Henry Ehrenreich, and David Turnbull. Solid State Physics. (Academic Press 1996). pp. 80–150.

[11] A. N. Morozovska, E. A. Eliseev, K. Kelley, Yu. M. Vysochanskii, S. V. Kalinin, and P. Maksymovych. Phenomenological description of bright domain walls in ferroelectric-antiferroelectric layered chalcogenides. Phys. Rev. B **102**, 174108 (2020).

[12] P. Guranich, V.Shusta, E.Gerzanich , A.Slivka, I.Kuritsa, O.Gomonnai. "Influence of hydrostatic pressure on the dielectric properties of $CuInP_2S_6$ and $CuInP_2Se_6$ layered crystals." Journal of Physics: Conference Series **79**, 012009 (2007).

[13] A. V. Shusta, A. G. Slivka, V. M. Kedylich, P. P. Guranich, V. S. Shusta, E. I. Gerzanich, I. P. Prits, Effect of uniaxial pressure on dielectric properties of $CuInP_2S_6$ crystals. Scientific Bulletin of Uzhhorod University. Physical series, **28**, 44 (2010).

[14]     A. Molnar, K. Glukhov, M. Medulych, D. Gal, H. Ban, Yu. Vysochanskii. The effect of changes in chemical composition and uniaxial compression on the phase transition of $CuInP_2S_6$ crystals, Abstract book of the FMNT 2020 Online Conference, Virtual Vilnius, Lithuania, 23 - 26 November (2020).

[15]     V. Maisonneuve, V. B. Cajipe, A. Simon, R. Von Der Muhll, and J. Ravez. Ferrielectric ordering in lamellar $CuInP_2S_6$. Phys. Rev. B **56**, 10860 (1997).

[16]     M. A. Susner, M. Chyasnavichyus, A. A. Puretzky, Q. He, B. S. Conner, Y. Ren, D. A. Cullen et al. Cation–Eutectic Transition via Sublattice Melting in $CuInP_2S_6$/$In_{4/3}P_2S_6$ van der Waals Layered Crystals. ACS Nano **11**, 7060 (2017).

[17]     J. Banys, J. Macutkevic, V. Samulionis, A. Brilingas & Yu. Vysochanskii, Dielectric and ultrasonic investigation of phase transition in $CuInP_2S_6$ crystals. Phase Transitions: A Multinational Journal **77:4**, 345 (2004).

[18]     V. Samulionis, J. Banys, Yu. Vysochanskii, and V. Cajipe. Elastic and electromechanical properties of new ferroelectric-semiconductor materials of $Sn_2P_2S_6$ family. Ferroelectrics **257:1**, 113 (2001).

[19]     A. Kohutych, V. Liubachko, V. Hryts, Yu. Shiposh, M. Kundria, M. Medulych, K. Glukhov, R. Yevych, and Yu. Vysochanskii. Phonon spectra and phase transitions in van der Waals ferroics MM'$P_2X_6$, Molecular Crystals and Liquid Crystals (2022). https://doi.org/10.1080/15421406.2022.2066787

[20]     https://www.wolfram.com/mathematica